\documentclass[lettersize,journal]{IEEEtran}

\usepackage{amssymb}
\usepackage[T1]{fontenc}
\usepackage[hyphens]{url}
\usepackage{booktabs}
\usepackage{caption}
\usepackage{graphicx}
\usepackage{xcolor}
\usepackage{float}
\usepackage{amsmath}

\usepackage{amsmath,amsfonts}
\usepackage{algorithmic}
\usepackage{algorithm}
\usepackage{array}
\usepackage[caption=false,font=normalsize,labelfont=sf,textfont=sf]{subfig}
\usepackage{textcomp}
\usepackage{stfloats}
\usepackage{url}
\usepackage{verbatim}
\usepackage{graphicx}
\usepackage{cite}
\usepackage{orcidlink}
\begin{document}

\title{Design and implementation of intelligent packet filtering in IoT microcontroller-based devices}

% \author[1,2]{Gabriel Victor C Fernandes%
%   %\fnref{fn1}
%   }
% %\ead{gabriel.fernandes@ga.ita.br}

% \author[1,2]{Pedro H Borges Monici%\fnref{fn2}
% }
% %\ead{pedroh.monici@ga.ita.br}

% \author[1,2]{César H de Araujo Guibo%\fnref{fn2}
% }
% %\ead{cesarguibo@usp.br}

% \author[1]{Gustavo de Carvalho Bertoli%\fnref{fn1,fn3}
% }
% %\ead{bertoli@ita.br}

% \author[3]{Aldri Santos%\fnref{fn1,fn3}
% }
% %\ead{aldri@dcc.ufmg.br}

% \author[1]{Lourenço Alves Pereira Jr.\corref{cor1}%\fnref{fn1,fn3}
% }
% %\ead{ljr@ita.br}

% \cortext[cor1]{Corresponding author | Postal address: Aeronautics Institute of Technology (ITA), Computer Science Division, Praça Mal. Eduardo Gomes, 50, 12228-900, S\~{a}o Jos\'{e} dos Campos, SP, Brasil | Email address: ljr@ita.br}

% \affiliation[1]{
% organization={Aeronautics Institute of Technology (ITA)},
% city={S\~{a}o Jose dos Campos},
% state={SP},
% country={Brazil}
% }

% \affiliation[2]{
% organization={University of S\~{a}o Paulo (USP)},
% city={S\~{a}o Carlos},
% state={SP},
% country={Brazil}}

% \affiliation[3]{organization={Universidade Federal de Minas Gerais (UFMG)},
% city={Belo Horizonte},
% state={MG}, 
% country={Brazil}}

\author{Gustavo de Carvalho Bertoli\orcidlink{0000-0003-1940-8295}, Gabriel Victor C. Fernandes\orcidlink{0009-0008-9402-7330}, Pedro H. Borges Monici\orcidlink{0000-0002-6959-0438}, César H. de Araujo Guibo\orcidlink{0000-0001-7748-0530}, Lourenço Alves Pereira Jr.\orcidlink{0000-0002-9682-0075}, Aldri Santos\orcidlink{0000-0002-5861-4414}

\thanks{Gustavo de Carvalho Bertoli and Lourenço Alves Pereira Jr. are with the Aeronautics Institute of Technology (ITA), S\~{a}o Jose dos Campos/SP, Brazil (e-mail: bertoli@ita.br, ljr@ita.br).}
\thanks{Gabriel Victor C. Fernandes, Pedro H. Borges Monici and César H. de Araujo Guibo are with the University of S\~{a}o Paulo (USP), S\~{a}o Carlos/SP, Brazil (e-mail: gabriel\_victor@usp.br, pedroh.monici@usp.br, cesarguibo@usp.br).}
\thanks{Aldri Santos is with Universidade Federal de Minas Gerais (UFMG), Belo Horizonte/MG, Brazil (e-mail: aldri@dc.ufmg.br).}
}

\markboth{Journal of \LaTeX\ Class Files,~Vol.~14, No.~8, August~2021}%
{Shell \MakeLowercase{\textit{et al.}}: A Sample Article Using IEEEtran.cls for IEEE Journals}


\maketitle

\begin{abstract}

Internet of Things (IoT) devices are increasingly pervasive and essential components in enabling new applications and services. However, their widespread use also exposes them to exploitable vulnerabilities and flaws that can lead to significant losses. In this context, ensuring robust cybersecurity measures is essential to protect IoT devices from malicious attacks. However, the current solutions that provide flexible policy specifications and higher security levels for IoT devices are scarce. To address this gap, we introduce T800, a low-resource packet filter that utilizes machine learning (ML) algorithms to classify packets in IoT devices. We present a detailed performance benchmarking framework and demonstrate T800's effectiveness on the ESP32 system-on-chip microcontroller and ESP-IDF framework. Our evaluation shows that T800 is an efficient solution that increases device computational capacity by excluding unsolicited malicious traffic from the processing pipeline. Additionally, T800 is adaptable to different systems and provides a well-documented performance evaluation strategy for security ML-based mechanisms on ESP32-based IoT systems. Our research contributes to improving the cybersecurity of resource-constrained IoT devices and provides a scalable, efficient solution that can be used to enhance the security of IoT systems.
\end{abstract}

\begin{IEEEkeywords}
packet filtering, intrusion detection, internet of things, constrained devices

\end{IEEEkeywords}

\section{Introduction}\label{sec:intro}

Cybersecurity is increasingly acting as one of the vital strategic aspects of business continuity~\cite{wef}. According to the World Economic Forum's 2021 Global Risk Report~\cite{wefrep2021}, incidents of this nature represent one of the most significant post-pandemic challenges. They potentially cause economic disruption, financial losses, geopolitical tensions, and social unrest. Thus, it is important to highlight that cybersecurity should be essential to the product and service development lifecycle. Although cyber attacks received attention mainly from specialized media in the past, the current status is different. Due to the digital transformation that the world is going through, these types of incidents have appeared in general media due to the disruption of services by cyber-attacks that affects the population directly~\cite{pipeline}.

High critical attacks can be achieved in this constantly changing environment by orchestrating large-scale compromised devices for malicious purposes. In this sense, a typical approach consists of a set of computational resources composing a command and control network (i.e., \textit{botnets})~\cite{iotbotnet,mantis}.  
Then, the attacker's objective is to compromise computers, smartphones, Wi-Fi routers, IP cameras, and others to compose this botnet. In this context, Internet of Things (IoT) devices are a common target because they usually have a precarious update, improper configuration, and maintenance procedures, as seen in attack campaigns like Mirai and Mozi~\cite{antonakakis2017understanding, mozi}. Moreover, this botnet risk tends to be present if the development practices do not ensure robust updating of security policies, intelligent security mechanisms, and secure-by-design. Additionally, it is common sense that IoT enables new technological, efficient, and profitable solutions. However, subverting IoT systems is profitable also to malicious actors~\cite{iotmarket,botnetmarket}. 

Nowadays cybersecurity mechanisms focusing on resource-constrained devices are scarce. More specifically, 
there is a lack of work that evaluates security mechanisms using machine learning for microcontroller-based IoT systems. Our work contributes to this research gap by proposing T800: a packet filter for Internet of Things (IoT) devices. T800 is a combination of mechanism and policy to implement a more secure operating environment for IoT systems, as it can work as an enabler to implement zero-trust architectures. The mechanism consists of instrumentation of the ESP-IDF framework TPC/IP stack, the lightweight IP (lwIP). It allows intercepting the network ingress traffic and deciding whether drop or not the current packet. The policies are the translation of machine-learning algorithms trained to identify malicious packets in the incoming network traffic. Hence, the mechanism allows the introduction of pluggable functions that returns boolean values to allow or not packets to proceed in the network stack. As contributions of this paper, we~highlight:

\begin{itemize}
\item Design and implementation of T800 allowing the deployment of rules for packet filtering with machine learning algorithms such as decision trees and neural networks in the ESP32 platform (FreeRTOS, TCP/IP protocol stack lwIP, and ESP-IDF SDK) with TensorFlow.  

\item A solution to allow IoT devices to be unnoticed during lateral movement or an internet-wide scanning campaign.

\item Performance evaluation of T800 indicating low overhead and energy efficiency. The results demonstrate reduction 
%improvements 
in resource consumption by removing unsolicited traffic from the protocol stack and avoiding processing them.
\end{itemize}

To the best of our knowledge, this is the first work evaluating the technical feasibility of machine learning-based detection on microcontroller-based systems. Further, the work is reproducible and open research for further developments in network security mechanisms for microcontroller-based systems. This paper is an extended version of a previously published paper in~\cite{sbrc}, and 
%In this work, 
we have enhanced the 
%previous 
work's scope and include new data and analysis to evaluate packet filtering models. Our results contribute to the knowledge advancement in packet filtering for security purposes and present important implications for the microcontroller-based systems and IoT domain.

We organize the remaining 
%sequence 
of the paper as follows:
in Section~\ref{sec:related}, related works are discussed. In Section~\ref{sec:t800}, we describe the architecture of the T800.
In Section~\ref{sec:method}, we describe the design of experiments. In Section~\ref{sec:result}, we report the results obtained and the influence of the factors. Finally,
Section~\ref{sec:conclusion} presents the conclusion of the  work 
%in Section~\ref{sec:conclusion}, 
and future~works.
%are listed.

\section{Related Works}\label{sec:related}

As observed in a broader application domain of computer networks and security, the characterization, identification, and creation of rules for packet filtering are well established~\cite{snort, signatures2021, signatures2021-2}. However, viewing them as componentized network functions to move them from the perimeter towards the endpoint of low computational power presents difficulties and restrictions~\cite{IMLADS}. Much is because these devices have limited functionality and high diversity in software systems. Furthermore, \cite{lifespan} showed that the model characterization solutions have a limited lifetime, meaning that there is evidence that attacks have time-varying behavior. The findings indicate a framed life of 6 (six) weeks, with an acceptable accuracy of 2 (two) to 8 (eight) weeks. Therefore, solutions that allow updating policies in response to new incidents are part of the requirements for sound performance.

Low-power devices constitute an essential part of the industrial scenario~\cite{idslowpower}. The reference shows an architecture to carry out the attack detection process in a distributed way. The focus is on the context of SCADA and PLC systems. As described, there is a precise characterization of the periodic communication behavior between the devices. Therefore, it allows the identification of anomalies. However, the study needs more generality, requiring an effort to implement new rules and apply them in other contexts.

In \cite{lightbulb}, the authors implement a microcontroller-based network intrusion detection system. They evaluate three different microcontroller architectures for their proposition. Their focus, however, consisted of loading the trained model on the resource-constrained device and sending network flow-based samples through a serial interface for only inference on the devices. Thus, this still needs to be a real security mechanism on a microcontroller-based system once an actual system would require to process the packet since its reception in the network interface.

Targeting a single-board computer platform, \cite{idsrasp} presents a solution for detecting malicious activities and provides results that indicate low computational demand for execution. Even though it is a flexible solution, the generalization of the filtering method and an update mechanism was outside the scope of the work. However, means for adapting to new attacks, revoking access, and incorporating new devices are out of their scope. In \cite{idsiot}, the authors implement a system focused on IoT devices using SVM (\emph{Support Vector Machine}) as an identification model. Even though the results indicate reproducibility, the data source consists of a conventional dataset without considering the implementation in an IoT device, with all conclusions based on MATLAB analysis. \cite{rnniot} presents neural networks as a potential approach. However, it does not discuss the need for a feasible update mechanism to tackle the evolutionary nature of network attacks. It neither implements nor evaluates its proposition on resource-constrained devices.

Our previous work~\cite{bertoli2021end} presents a framework that describes the creation of packet filtering policies. It is a methodology that helps in the characterization process of known attacks, allowing the use of machine learning models to prevent them. A vital contribution deals with implementing the model in a real environment so that the numerical results become comparable with the feasibility of deploying the models and the performance in a target system during execution. However, the systematization of a solution to integrate with low-power platforms (i.e., microcontrollers) is not considered. Additionally, this work discusses updated mechanisms due to the inherent characteristics of traffic changes (i.e., concept drift). However, compared with the approach in this work, it is not through a modular update in the model loaded in the software implementation but a new execution of the whole framework.

\cite{kitsune,passban, iotkeeper} presents an anomaly-based intrusion detection system. They use real traffic generated by testbeds composed of IoT devices/sensors to evaluate their proposition. However, in contrast to our approach, these propositions consider the intrusion detection system part of a gateway instead of deploying the solution in the node device. In all cases, no update mechanism is discussed, and their runtime performance evaluation on a single-board computer considers only network throughput for \cite{kitsune}. \cite{passban,iotkeeper} evaluates metrics such as CPU, network bandwidth, and memory consumption but no power consumption. 

In this work, we design, implement, and evaluate machine learning algorithms in an ESP32 ($320$\,kB of RAM) for protecting IoT devices (i.e., edge nodes), demonstrating adequate performance. Another differentiation from our work compared to the knowledge on intrusion detection for IoT is the lack of low-level implementation (i.e., kernel mode) for the security mechanism. Usually, they work with user-space implementations and commonly consider the use-case of deployment in an IoT gateway. In our case, we use edge deployment in the device itself. Additionally, works consider more powerful devices than microcontroller-based systems, such as single-board computers like Raspberry Pi~\cite{kitsune,passban,iotkeeper,8991929,bertoli2021end}. We evaluate and propose a low-level implementation tied to the network stack on microcontrollers (Lightweight IP -- lwIP).

We summarize all the presented related works in Table~\ref{tab:related-works}. Our findings show IoT, IT, and OT domain applications. Next, we evaluate the presence of an update mechanism for the security mechanism, for this attribute only our work presents an specific solution. We are also assessing the deployment of T800 on resource-constrained devices, the microcontroller-based system. Regarding runtime performance evaluation, our scope is CPU, memory, network, and in addition to previous works, the power consumption of the proposed mechanisms.

Most related works focus on point methods with restricted generality and often lack on-device deployment (edge nodes). T800 differs from them because it has a design adaptable to other platforms with low computational performance, capable of coupling to different TCP/IP protocols and operating systems stacks. Our solution also allows IoT devices to drop malicious traffic, increasing resource-constrained devices' efficiency. Finally, it demonstrates the effectiveness of implementing machine learning filtering policies. T800 enables the execution of the decision trees, logistic regression, SVM, and multilayer perceptron algorithms. Hence, the present work advances state-of-the-art by making security policies directed to packet filtering viable in resource-constrained IoT devices.

% \usepackage{booktabs}
% \usepackage{graphicx}
\begin{table*}[]
\centering
\caption{Related works for IoT attack detection / security-mechanisms.  The  \checkmark\, means the presence of the attribute under evaluation, in the resource-constrained device the evaluated microcontroller is presented. For runtime performance evaluation, the complete scope considers CPU, Memory, Network and Power.}
\label{tab:related-works}
\resizebox{\textwidth}{!}{%
\begin{tabular}{@{}ccccc@{}}
\toprule
 & \textbf{Domain} & \textbf{Update Mechanism} & \textbf{\begin{tabular}[c]{@{}c@{}}Resource-constrained\\ Device\end{tabular}} & \textbf{\begin{tabular}[c]{@{}c@{}}Runtime Performance\\ Evaluation\end{tabular}} \\ \midrule
\cite{idslowpower} & OT &  & \checkmark \,\,(ARM Cortex-M7) & only network \\
\cite{lightbulb} & IoT &  & \checkmark \,\,(ESP32, ESP8266, ATMega328p) &  \\
\cite{idsrasp} & IoT &  &  &  \\
\cite{idsiot} & IoT &  &  &  \\
\cite{rnniot} & IoT &  &  &  \\
\cite{bertoli2021end} & IT &  &  & \checkmark (no power) \\
\cite{kitsune} & IoT &  &  & only network \\
\cite{passban} & IoT &  &  & \checkmark (no power) \\
\cite{iotkeeper} & IoT &  &  & \checkmark (no power) \\
\cite{8991929} & IoT &  &  &  \\
\textbf{Our work} & IoT & \checkmark & \checkmark\,\,(ESP32) & \checkmark \\ \bottomrule
\end{tabular}%
}
\end{table*}

\section{T800 --- Packet filtering}\label{sec:t800}

We evaluates the design, implementation, and evaluation of packet filtering in an ESP32 microcontroller ($320$\,kB of RAM), aiming at a low-computational cost due to its targeting to embedded devices. Thus, our design of a system aimed at edge devices must address specific requirements to ensure the feasibility of running on the target computing platform. These requirements involve coupling the underlying software system with low overhead when running. Further, we use machine learning-based security policies jointly
%in conjunction 
with a zero-trust architecture, considering the extensive adoption of IoT solutions with constantly evolving workloads.

In this context, the T800 project enables the implementation of security policies with such requirements. 
%The present proposal 
T800
is limited to the \textit{stateless} packet filtering functionality, but 
%However, 
it is possible to implement other models 
taking into account
%considering 
\textit{stateful} attributes or anomaly detection~\cite{flids}. Thus, 
%the present work 
we implement an intelligent packet filter suitable for embedded devices. More specifically, the T800 uses ESP-IDF~\footnote{esp-idf: \url{https://github.com/espressif/esp-idf}}, an IoT open-source framework developed by Espressif, aiming at implementing user application for ESP32 platform. Therefore, T800 packet filter follows a design that promotes low computational consumption and, simultaneously, a generic enough structure to allow expansion to other systems.

\subsection{Architecture}
\begin{figure*}[h]
    \centering
    \includegraphics[width=\textwidth]{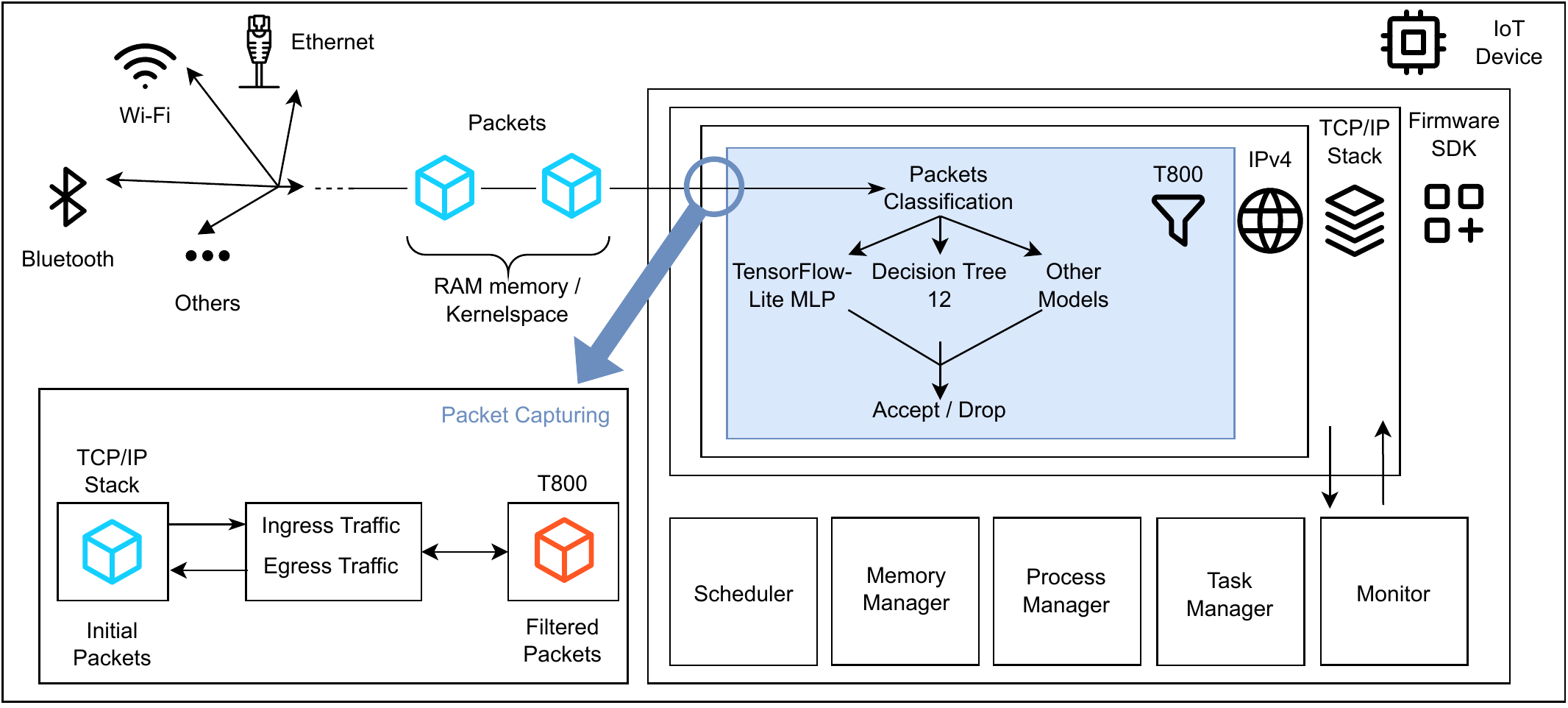}
    \caption{T800 packet filtering architecture.  T800's mechanism consists of intercepting the network packets after they are available in RAM.  The interception point corresponds to the first function inside the TCP/IP stack to ensure filtering before processing by the adjacent layers.  The decision to accept or drop packets is pluggable, leaving the security policy specification flexible and dynamic.}\label{fig:arquitetura}
\end{figure*}

The T800 component, illustrated in Figure~\ref{fig:arquitetura}, makes 
%implements 
an evaluation of network traffic received by an ESP32-based device. First, it evaluates the packet header to differentiate the traffic as malicious or benign. 
Thus, T800 captures each network packet entering the TCP/IP stack. The implementation follows the function responsible for processing the TCP/IP protocol stack packets within the Lightweight~IP (lwIP) component in ESP-IDF.

For this purpose, T800 was built as a new component of the framework ESP-IDF, becoming part of the standard library. Additionally, it requires two new dependencies: \textit{esp-nn}\footnote{esp-nn: \url{https://github.com/espressif/esp-nn}}, an official Espressif library that implements common functions for assembly-optimized machine learning; and \textit{tflite-lib}\footnote{Tensorflow Lite: \url{https://github.com/tensorflow/tflite-micro}}, a Google library that makes possible to deploy Machine Learning models developed in Tensorflow on ESP32. \textit{tflite-lib} implements optimizations for porting conventional models to devices with low-computational resources, through techniques such as quantization of the weights of the final models.

While running on the ESP32, T800 undergoes an initialization step.  This initialization step takes place by providing an initial configuration structure, including a function that classifies packets and information about the function execution context (static or dynamic).  This type of context is necessary because T800 can operate in two different ways, whether storing the TCP flow state (\textit{stateful}) or not (\textit{stateless}).  During the T800 initialization, it runs in the foreground of the process in a dedicated thread (i.e., FreeRTOS task).  As a result, FreeRTOS preempt the ready processes (tasks) among the available CPUs, and T800 is one of them.  After the initialization step, T800 starts to act directly on the Network and Transport layers, capturing the packets provided by one of the network interfaces.  The function that uses a machine learning model to classify packets is chosen and defined in advance in the initial configuration structure.  The input consists of data from the TCP and IP headers.  These data are present in the packets received by ESP32 through a structure called \texttt{pbuf}, representing a packet in the lwIP stack~\cite{dunkels2001design}.  It contains the headers for TCP/IP, the data link layer, payload data, and, when necessary, a reference to other packets that may be part of a sequence.

Finally, the lwIP stack may or may not process the packet. If the model output is malicious, T800 discards the current \texttt{pbuf}. On the other hand, if it classifies as a non-malicious packet, the packet follows its processing in ESP-IDF. Our implementation is limited to the Ingress Traffic provided by one of the network interfaces, meaning T800 can avoid any incoming malicious traffic entering the ESP32-based IoT device.

\section{Methodology}\label{sec:method}

This section presents the T800 modular implementation and its update mechanism. Next, we detail the creation of the packet filtering rules with machine learning and describe the port scanning use case under consideration in this work. We discuss the dataset used for model training and the machine learning parameters. Later, we present the thorough computational metrics evaluation methodology.

\subsection{T800 Implementation}

\begin{figure}[htb]
    \centering
    \includegraphics[width=\columnwidth]{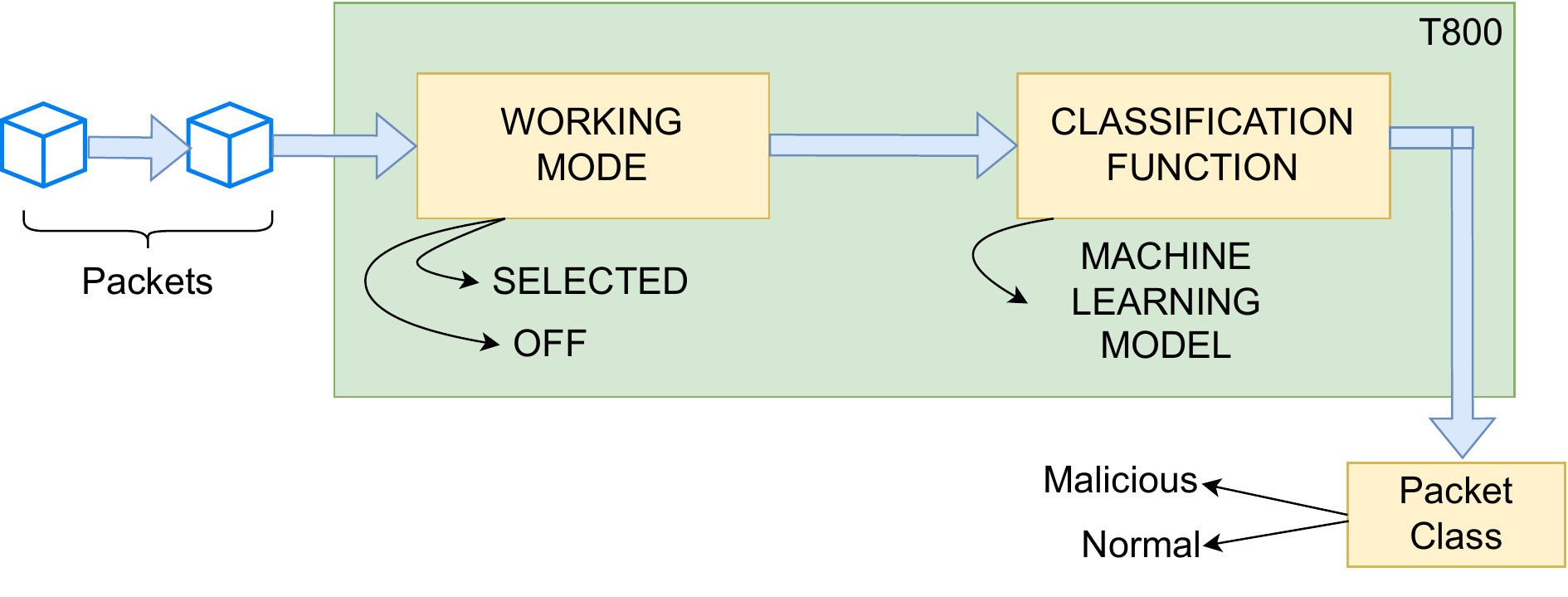}
    \caption{Implementation structure of T800.  The current packet workflow is dynamically activated or deactivated depending on the execution context.  In addition, the pluggable classification function implements the heuristics to identify malicious or normal packets.}\label{fig:estrutura}
\end{figure}

The implementation of T800 has the following main objectives: low-computational cost and the ability to update filtering policies (\emph{update mechanism}). In this context, the filter implementation establishes a standardized interface that, with few resources, facilitates the development and implementation of new filtering rules. Figure~\ref{fig:estrutura} depict this structure. It has only two essential entities: a value responsible for encoding how it works (working mode) and a classification function that enforces a packet filtering policy (classification function). The working mode defines how the system performs the packet capture and when it executes the classification function, including its arguments. The classification function receives the required context for performing the classification and returns the classification of the packet. Therefore, given requirements for execution time and memory footprint, the possible choices could be chosen to achieve the most suitable working mode and the best policy available. An important attribute is that this approach allows the alteration of both entities at runtime, thus providing greater flexibility to the filter, achieving the update mechanism requirement.

\subsection{Packet filtering rules}

A packet filtering policy specifies whether a packet continues its natural processing path within a protocol layer. Thus, considering how general this definition may be, there are numerous ways to construct such a policy. Its construction can range from rules selected by a domain expert to unsupervised machine-learning models.

T800 follows the AB-TRAP framework~\cite{bertoli2021end} that seeks to design protection mechanisms based on machine learning models. AB-TRAP covers the whole development chain, from the study of the normal behavior characterization to the implementation and performance evaluation of the proposed solution to the operation. Thus, the packet filtering solutions tested on the T800 followed this same framework. First, they correspond to generating data from the attacks and training the models on this data. Then, we implement it on the ESP32 device for performance evaluation. Finally, the filtering policies correspond to machine learning models trained offline, which may require periodic updates according to the needs of each application and environment.

In this context, we employ five different filtering policies to test the operational viability of a low computational cost platform. Another point considered was to verify the applicability of the T800 in preserving the ability to generalize the rules. Therefore, one policy with a Decision Tree (DT) structure with depth 12 (DT-12) is under analysis. Meanwhile, a Multilayer Perceptron (MLP) with two hidden layers with 16 neurons each and an output layer with two neurons is under evaluation, as well as a Logistic Regression (LR) and a Support Vector Machine (SVM) with a linear kernel. On the other hand, we consider the standard system as baseline (i.e., without T800); this approach allows us to verify how our solution impacts the system resources. Decision Trees require less computing power when put into operation, whereas Multilayer Perceptrons are more intensive than the former.  Thus, choosing them to compare allows us to explore a wide range of classical machine learning algorithms in the computing exigency spectrum.  A key point is the algorithm implementation feasibility in our testbed, the Espressif ESP32.  Therefore, we intend to verify the on-device cost of models with both high and low expected computing costs.

%% Thus, these models fit well in implementing the \textit{T800} component since minimizing the computational cost in this type of system is essential due to the scarcity of resources. Similarly, the Logistic Regression and the SVM models exhibit the same qualities, considering the system's limited resources. Furthermore, these models were implemented through the Tensorflow framework and then converted to the final model of ESP32 through Tensorflow-Lite.

The Decision Trees (DT) correspond to a chain of conditional structures based on attribute tests derived from the original models. Interpretability and portability (from training to implementation) are key features that make DT feasible to low-power computing platforms, requiring low memory footprint and computing power. Thus, these models are a good fit to implement as policies for the T800 component. Similarly, the Logistic Regression and the SVM models exhibit the same qualities concerning the system's limited resources. Furthermore, these models were implemented through the Tensorflow framework and then converted to the final model of ESP32 through Tensorflow-Lite.

The MLP is specified with a sigmoid activation function for the hidden layers and a softmax activation function for the output layer. Both its implementation and training are done with Tensorflow and converted to the final model on ESP32 with Tensorflow-Lite. This conversion makes this model adopt space and processing optimizations that offer satisfactory performance in implementing inference mechanisms on embedded and resource-constrained devices.

\subsection{Use-case, dataset and model training}

%Studies carried out on $7.41$ Petabytes of traffic data, collected from a large {\em backbone} network between 2004 and 2011, showed that about $2.1\%$ of packets are  {\em scan} \cite{Glatz:2012:CIO:2398776.2398781}.  This data reveals the dimension that these attacks represent on the Internet.  Indeed, the annual traffic has doubled every two years since 2005, reaching $2.3$ Zettabytes today, also driven by the deployment of 5G and the growth of the Internet of Things devices~\cite{8385276, Lee:2012:TSI:2427036.2427038,iot_forecast}.  %However, solutions focused on smaller devices are scarce.  
In this work, the detection of port scanning attacks is the use-case under analysis. Port scanning is part of the reconaissance step of an attack, the reconaissance is the first step according to the cyber kill chain framework~\cite{ckc}. Thus, avoiding an attack on its initial phase, saves resources and reduces its possible impact. Considering the IoT networking characteristics, most attacks relies on scanning the networks to identify visible and vulnerable IoT devices.  %Solutions to allow the implementation of specialized security policies based on \textit{zero-trust}~\cite{zerotrust,zta} mitigate the risks in upper layers.
Also, scanning the local infrastructure (e.g., lateral movement) or throughout the Internet, thus port scanning represents a critical attack vector to IoT systems~\cite{ durumeric2014matter, durumeric2015search, ros, tls}. 

Regarding the model training, the decision tree uses the entropy metric for the training stage as a division criterion. The MLP was trained for $2,000$ \textit{epochs} with the Adam optimizer, a learning rate of $1.10^{-5}$, and a batch size of 260.

The dataset for all training steps is the AB-TRAP~\cite{bertoli2021end} with a small change that consists of removing the \texttt{tcp.window\_size} attribute. This dataset generates attack packets through a testbed and a collection of exclusively port scanning malicious traffic. For this, we used TCP scanners such as \texttt{Zmap}, \texttt{masscan}, \texttt{Hping3}, \texttt{Unicorn Scan}, and \texttt{NMap}, resulting in 86.480 malicious packets. On the other hand, MAWILab \cite{fontugne2010mawilab}) to represent the benign samples. The dataset consists of $103,094$ packets sampled on November 21, 2019, from an Internet link that connects the United States of America to Japan. After that, the malicious and benign traffics are merged into a single dataset through a \textit{salting} approach. From this process, the results of \textit{F1-score} 0.79, 0.34, 0.93, and 0.91 were obtained for the Logistic Regression, SVM, Decision Tree, and the MLP, respectively.

\subsection{Computational Metrics Evaluation}\label{sec:eval}

%This section presents the methodology for evaluation and performance analysis of the \textit{T800}. 
The T800's performance is measured using four metrics summarized in Table~\ref{tab:metrics}. The first is the CPU utilization rate of the two cores present in ESP32. The second is the amount of memory allocated by the T800 (only static memory is measured as the component does not perform dynamic memory allocation). The third is the rate at which the Wi-Fi interface receives packets. The last one is the power consumption of the device. These metrics are similar to the performance, memory, and energy metrics proposed by \cite{benchiot}.

\begin{table}[h]
    \caption{Measurements collected and evaluated for T800.}
    \centering
    \resizebox{\columnwidth}{!}{%
    \begin{tabular}{lll}
    \toprule
    \textbf{Metric} & \textbf{Description}\\
    \midrule
    CPU & CPU usage (all cores)\\
    MEM & Memory usage (stack only)\\
    NET & Network usage (Wi-Fi only)\\
    POWER & Energy consumption in milliwatts (mW)\\
    \bottomrule
    \label{tab:metrics}
\end{tabular}
}
\end{table}

Software developed for edge IoT devices usually works in the context of scarce computing resources. Thus, these metrics are relevant to the context of embedded systems, which is why we consider them in our analysis. For example, the ESP32 has only $320$\,kB of RAM and a dual-core processor with a clock cycle of $240$\,MHz. Thus, verifying eventual increases in the lwIP processing rate or allocated memory is mandatory to ensure the system's operation. Furthermore, the system may become overloaded depending on the packet rate experienced by the network interface due to such computing constraints. Thus, monitoring the network utilization is crucial to understanding how T800 impacts the TCP/IP protocol stack implementation of lwIP. Finally, as some IoT devices could be battery-powered, it is also necessary to understand the impact of the T800 on energy consumption.

Based on these metrics, we consider different execution environments to contrast the ESP32 operation with and without the T800 in different situations. Table~\ref{tab:design} describes the design of experiments we adopt in our study. Two factors are under consideration for simulating different execution scenarios. One of them is the intensity of benign network traffic received on the Wi-Fi interface (I), which can be $8$\,Mbps (I0) or $16$\,Mbps (I1). The other is the presence of malicious packets in the network traffic destined for ESP32 that varies between the values absent (M0) and present (M1). We refer to the experiments as IxMy codes in all possible combinations, \{M0I0, M0I1, M1I0, M1I1\}.

\begin{table}[h]
    \caption{Properties of the traffic during the experiments.}
    \centering
    \resizebox{\columnwidth}{!}{%
    \begin{tabular}{lll}
    \toprule
    \textbf{Property} & \textbf{Level} & \textbf{Code}\\
    \midrule
    Traffic intensity & 8\,Mbps or 16\,Mbps & I0, I1\\
    Malicious traffic & Absent or Present & M0, M1\\
    \bottomrule
    \label{tab:design}
\end{tabular}
}
\end{table}

\begin{figure}[htb]
    \centering
    \includegraphics[width=\columnwidth]{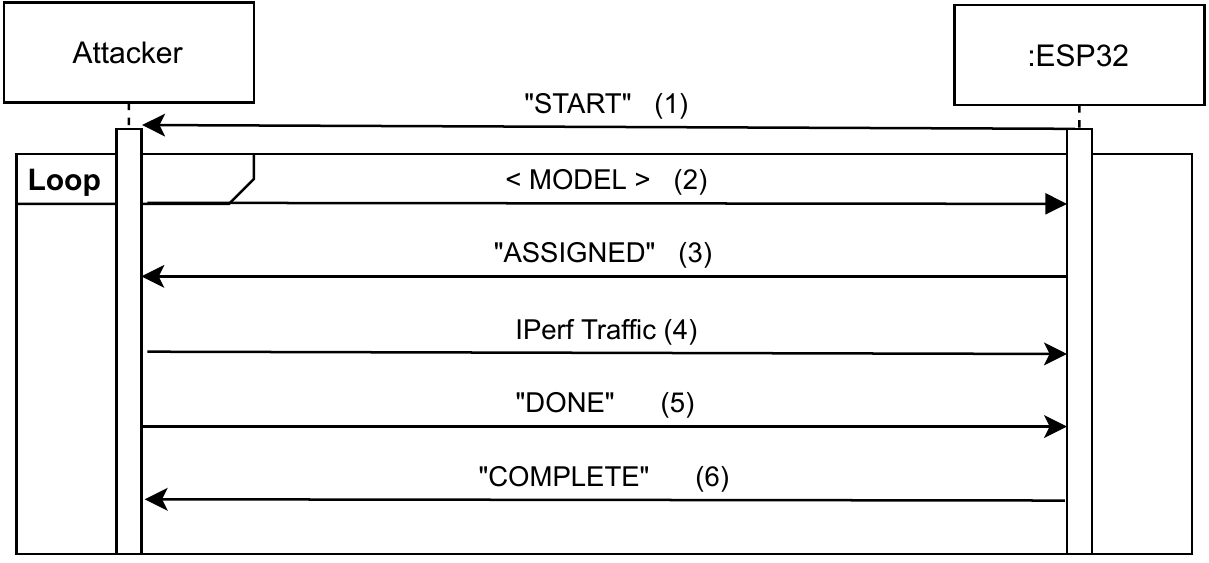}
    \caption{Sequence diagram for ESP32's evaluation metrics capturing.  Describe the protocol to execute an experiment replica.  It dynamically loads the corresponding classification function, and the workload excites during a period after the system signaling is ready.}
    \label{fig:dss}
\end{figure}

The messages between the test devices occur between an ESP32 and an attacking computer on the same wireless network. The attacking computer is responsible to generate both benign and attack traffic (over TCP), as well as, collect all the experiment data (over UDP). To do this, they communicate through the UDP protocol to manage the settings of a TCP connection that will be active for $360$ seconds. As depicted in Figure~\ref{fig:dss}, in the first execution step (1), ESP32 sends a message to the attacking machine signaling the beginning of the experiment. In the second (2), the attacker sends a code that can specify the T800's filtering policy or disabling T800. Then, in the third step (3), the ESP32 responds by communicating that it has received the necessary information and has already completed its initial configuration. In this process, two servers start on ESP32. The first receives all TCP traffic from the simulation. In contrast, the second collects performance metrics and sends them to the attacking machine via UDP at a 1-second rate. After that, in the fourth step (4), the network traffic is generated by the attacking machine. Finally, after collecting all the metrics, in the fifth step (5), the attacking machine sends a message to ESP32 that represents the end of the experiment. Finally, in the sixth step (6), the ESP32 sends a response confirming the end of the experiment. It is worth to note this approach allows the packet filtering policy to be changed at runtime (step 2), enabling the solution's adaptability.

\begin{figure}[h]
    \centering
    \includegraphics[width=.75\columnwidth]{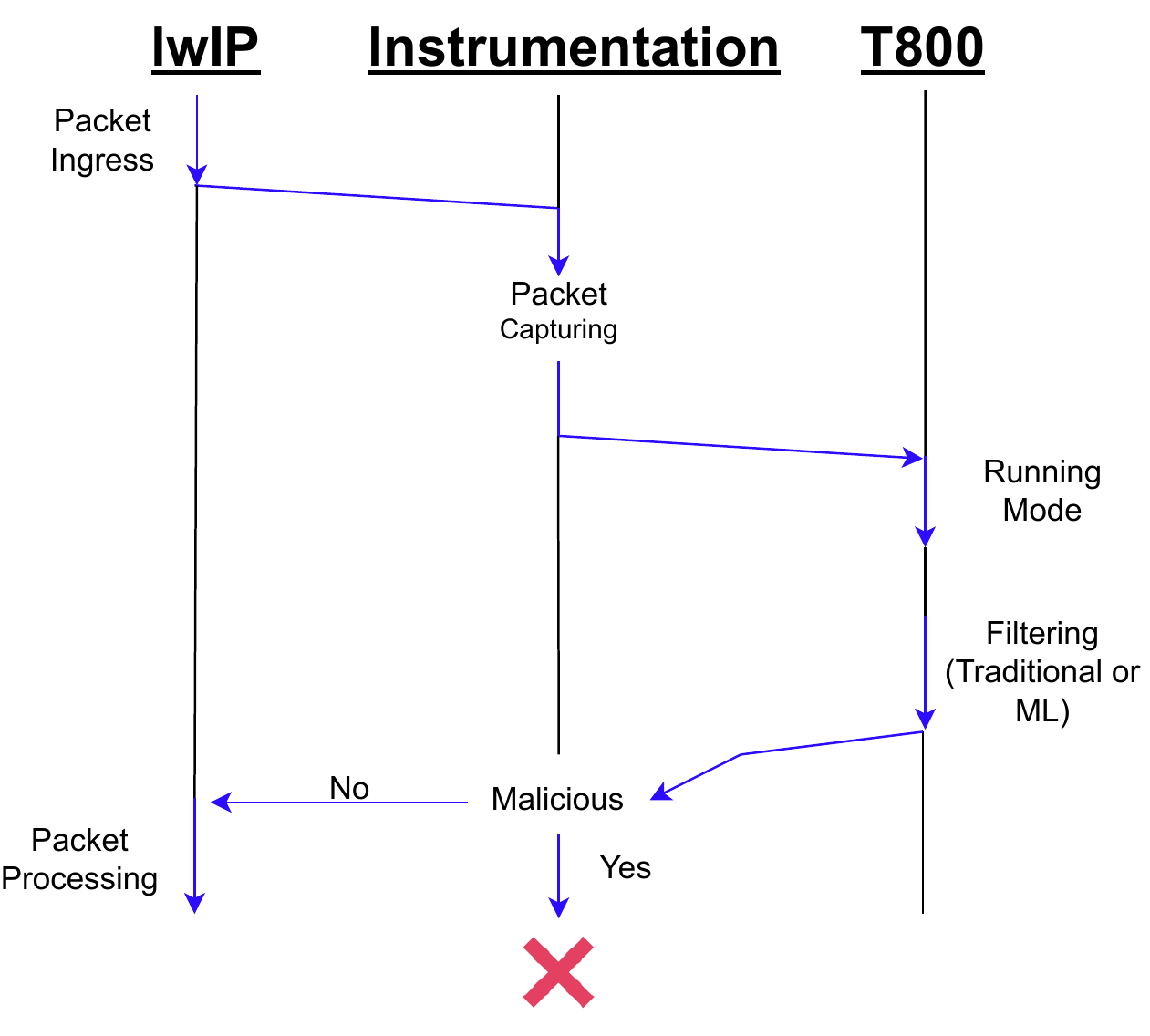}
    \caption{Interaction of T800 with ESP-IDF stack lwIP.  T800's design assumes coupling to the lwIP, influencing the minimum with the original system.  However, it is flexible to adapt to other stacks.}\label{fig:interaction}
\end{figure}

As illustrated in Figure~\ref{fig:interaction}, the packet filter integrates with the system's base TCP/IP stack (lwIP). First, packets ingress through the stack and are intercepted by the instrumentation part before reaching T800. Then, T800 is configured with its working mode to perform the filtering. Such filtering can be traditional or advanced. The first performs static packet filtering rules without requiring complex algorithms. The advanced one follows the approach taken in our experiments with the help of machine learning models. Finally, the packet can be classified as malicious or not and can be processed in lwIP or discarded. This integration allows T800 to have an in-depth view of the system, which allows the measurement of computational cost.

\begin{figure}[htb]
    \centering
    \includegraphics[width=\columnwidth]{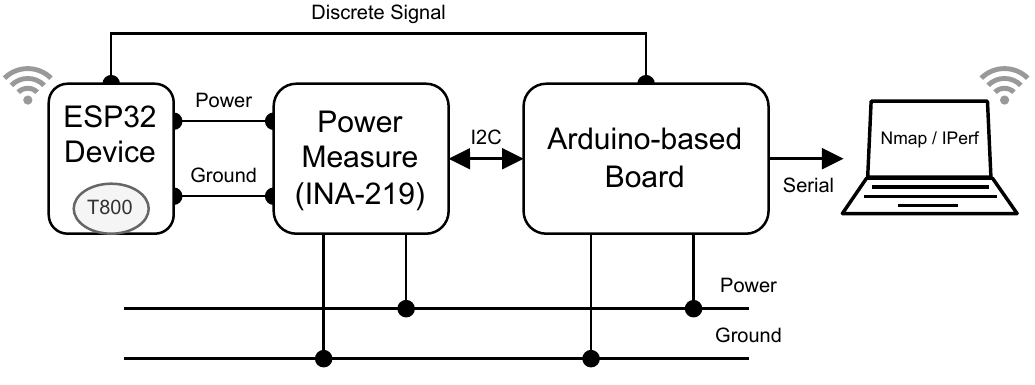}
    \caption{Setup for measuring the T800 energy consumption in an ESP32-based system.  The monitor is implemented in hardware and attached to the physical system.  All data gathered is stored separately on the monitor.}\label{fig:pwr}
\end{figure}

All the performance metrics except by the energy consumption are available from the FreeRTOS API functions implemented in ESP32. For power consumption, the ESP32 power pins with a current/power sensor\footnote{Texas Instrument INA219: \url{https://www.ti.com/lit/ds/symlink/ina219.pdf}} allows to monitor the energy consumption by hardware. The sensor transmits the readings of these measurements by I2C communication to a reading device (Arduino-based) that uses serial communication with the test computer, in our case the same as the attacker, to record these measurements (serial over a USB connection). In conjunction with the energy measurements, the reading device also monitors a discrete signal from ESP32 that is responsible for indicating the start and conclusion of an experiment, thus facilitating the post-processing of the data. In addition, the
normal traffic with intensities of $8$\,Mbps (I0) and $16$\,Mbps (I1) are generated through \textit{IPerf
v2.0} and malicious traffic is generated with \textit{Nmap} to perform scanning attacks both through wireless communication. Figure~\ref{fig:pwr} depicts the setup.

This work adopt a complete factorial design of experiments. Experiments I0M0, I0M1, I1M0, and I1M1 runs with all filtering policies and without T800. Further, we collect all measurements in a thirty ($30$) replicas experiment execution. Finally, the hardware specifications of both the test/attacker computer and the ESP32 in use are present in Table~\ref{tab:hardware}, along with those of the device used to measure energy consumption during collection (later test computer). 

\begin{table}[t]
 \centering
 \caption{Description of the hardware used in the experiments.}
 \label{tab:hardware}
\resizebox{\columnwidth}{!}{%
    \begin{tabular}{llrl}
        \toprule
        \textbf{Description} & \textbf{CPU (GHz)} & \textbf{Memory} & \textbf{Operating System}\\
        \midrule
        Test computer & Intel i7-8565U-(4.6) x8 & 16\,GiB & Manjaro Linux x86\_64\\
        ESP32 DevKit V1 & Xtensa-(0.240) x2 & 320\,kB & FreeRTOS, ESP-IDF\\
        Test computer & Intel i5-3337U-(2.7) x4 & 6\,GiB & Ubuntu 20.04.2 LTS\\
        \bottomrule
    \end{tabular}%
}
\end{table}

\section{Results and Discussions}\label{sec:result}

\begin{figure}[h]%{\columnwidth}
    \centering
        \includegraphics[width=0.5\columnwidth]{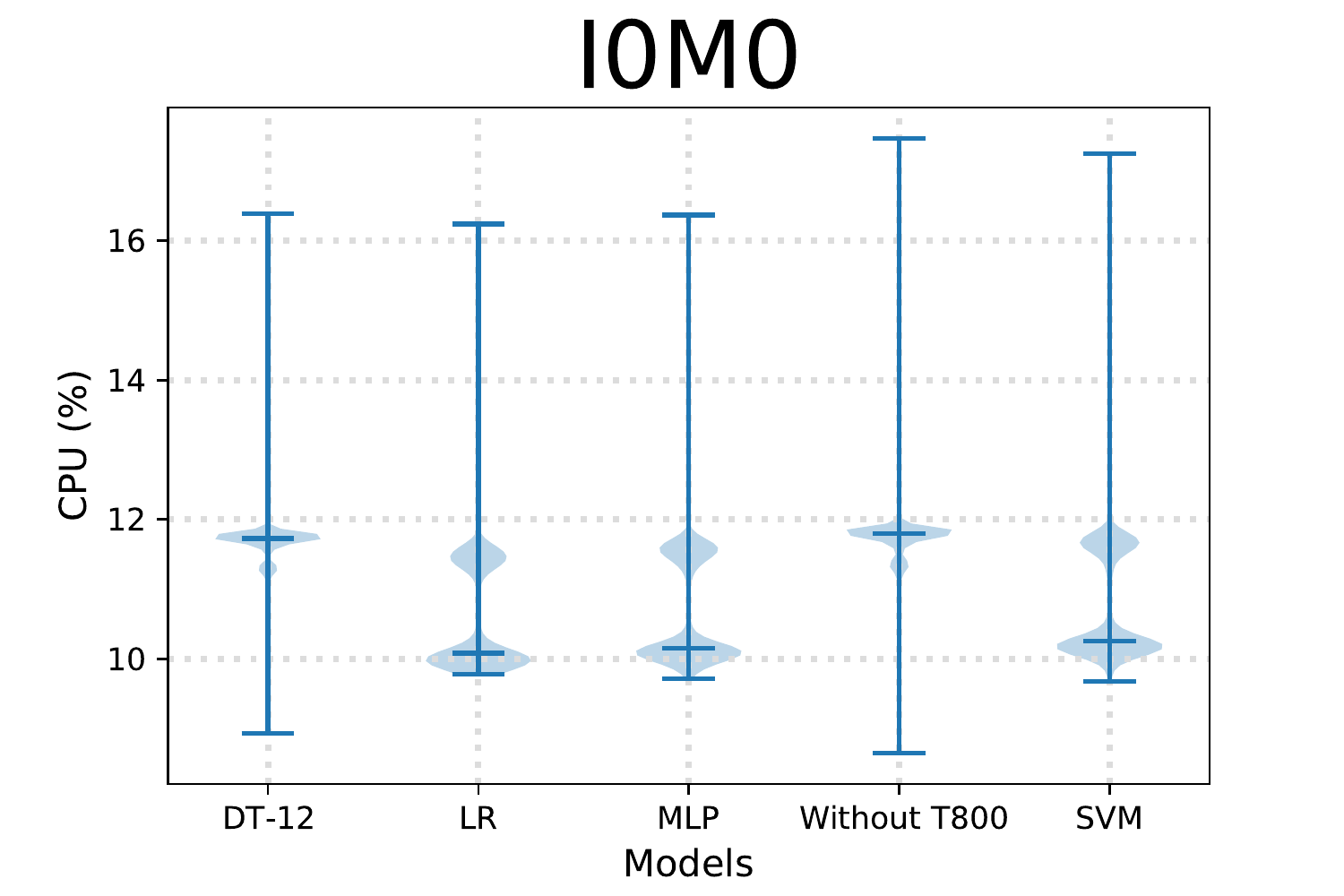}\hfill
        \includegraphics[width=0.5\columnwidth]{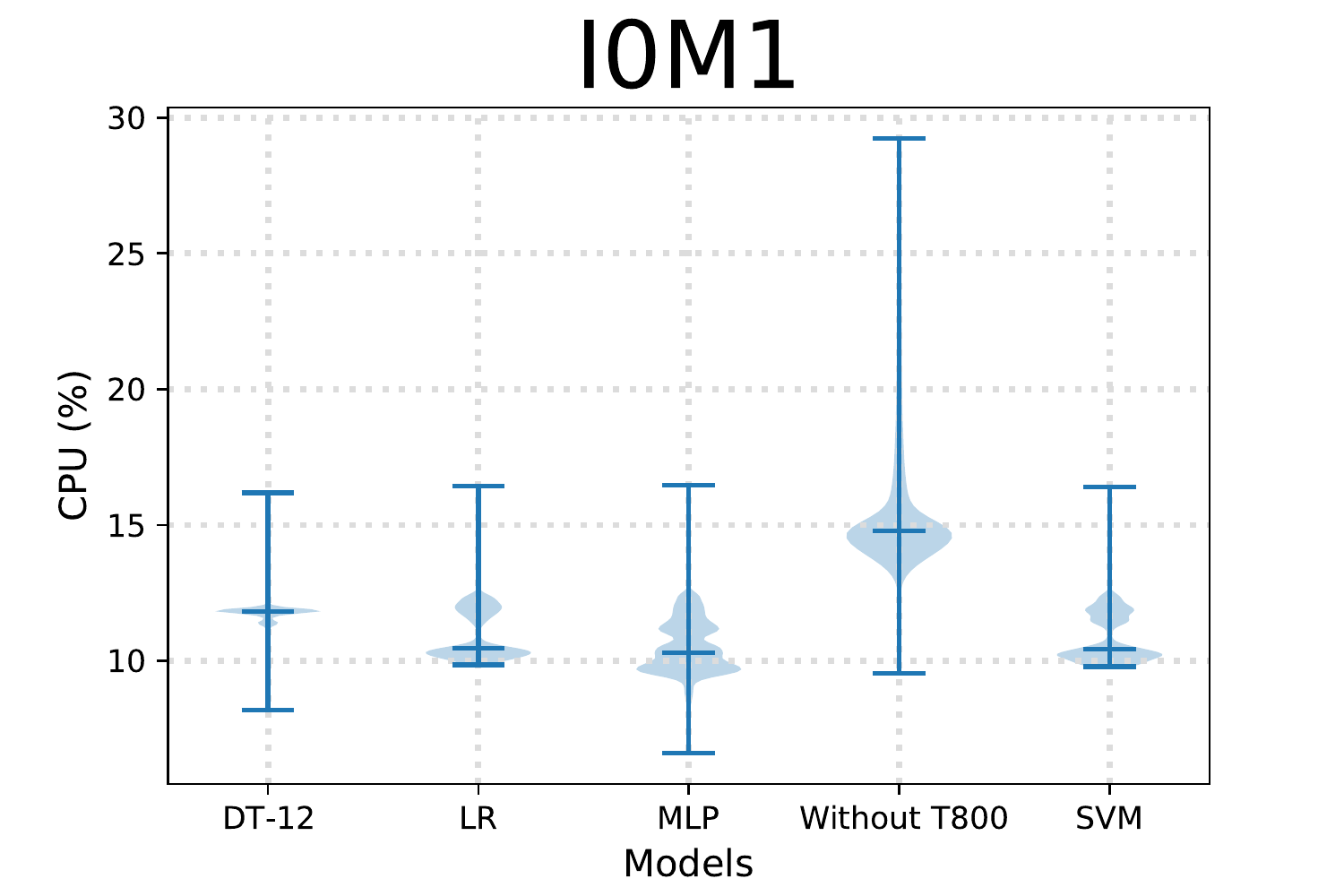}
        \includegraphics[width=0.5\columnwidth]{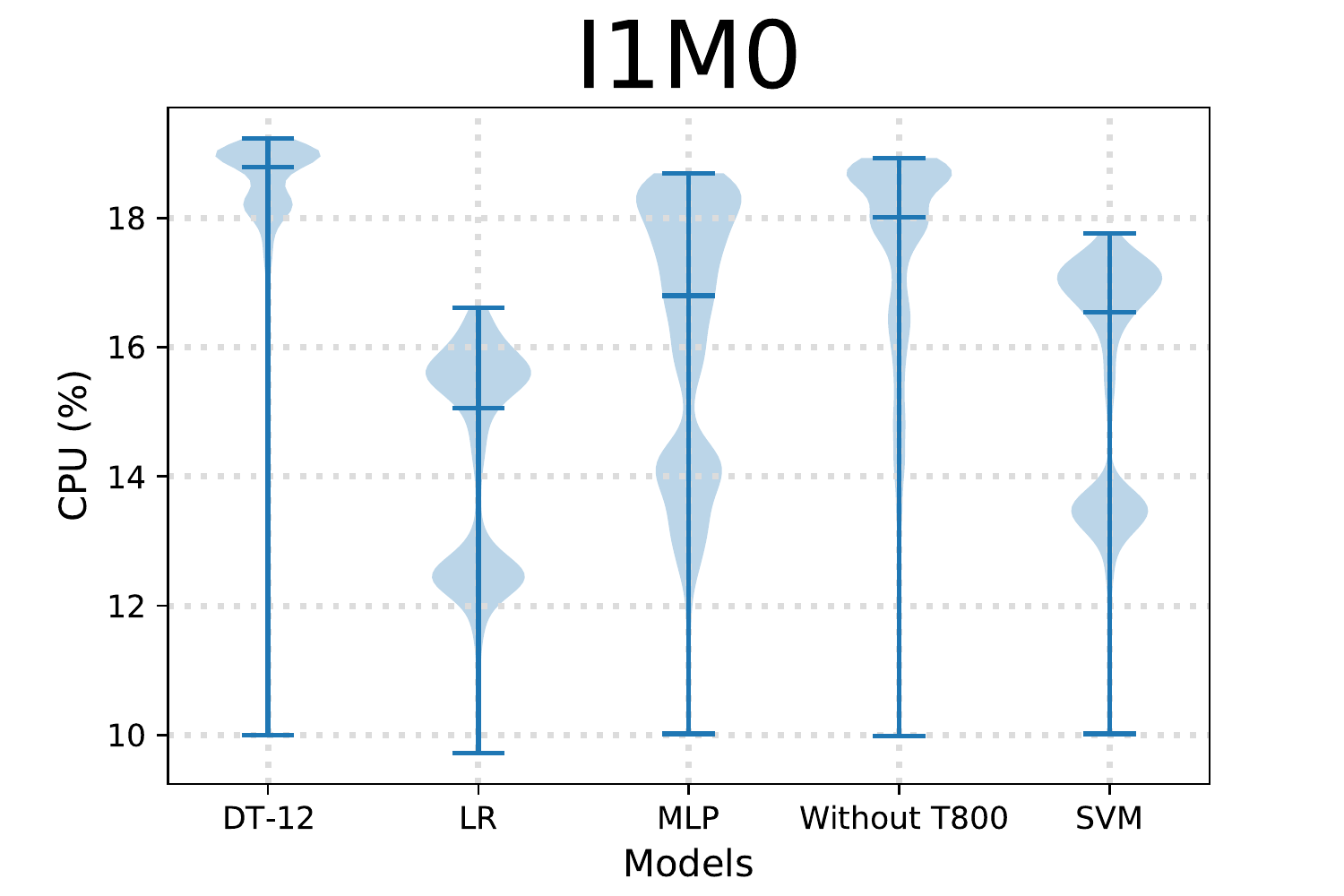}\hfill
        \includegraphics[width=0.5\columnwidth]{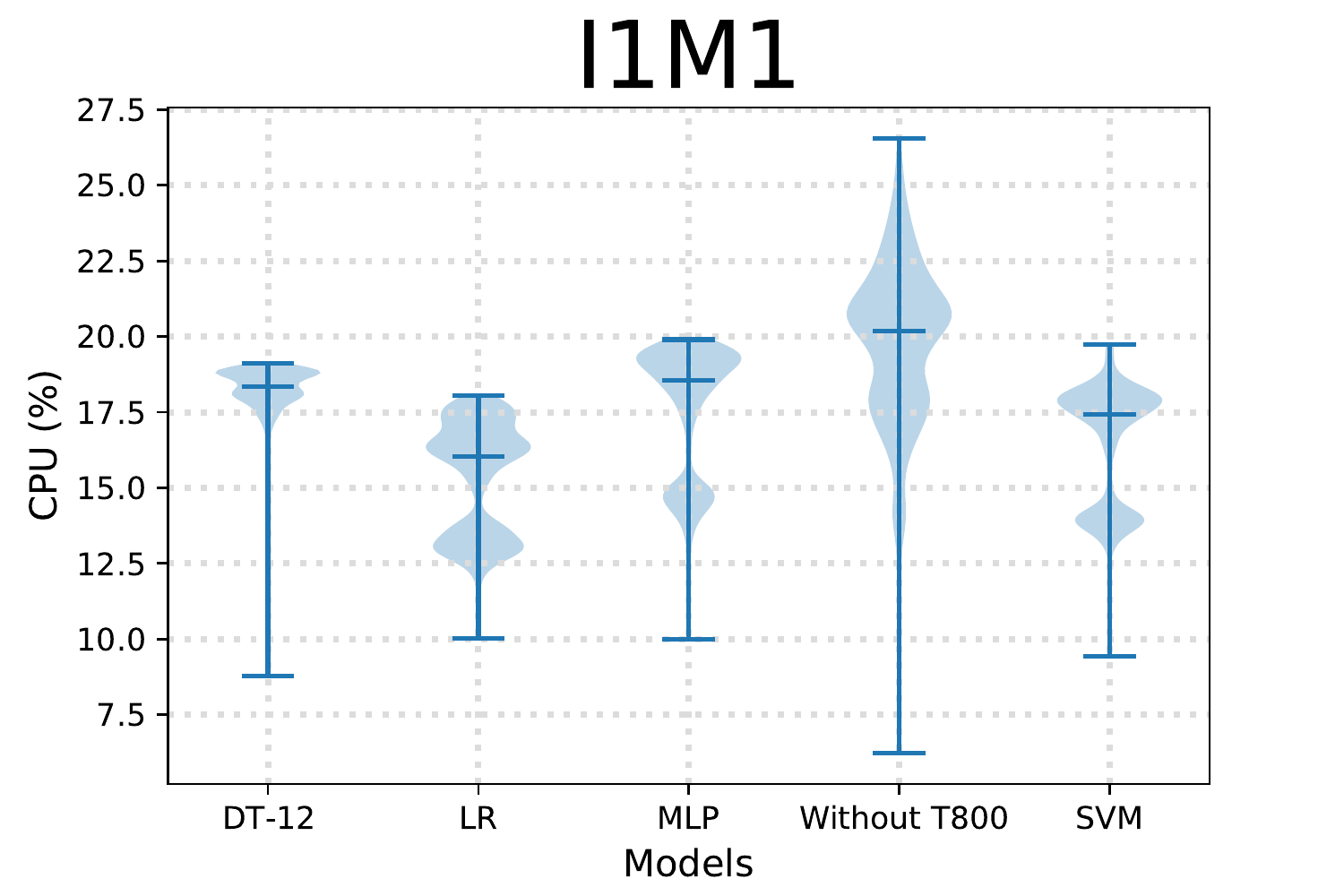}
        \caption{Consumption values obtained from experiments: CPU Usage.  The row I0Mx corresponds to low-intensity traffic ($8$\,Mbps) and I1Mx to the high intensity ($16$\,Mbps).  The column IxM0 indicates the absence of malicious traffics, and IxM1 the presence.  T800 enables the system to experience lower CPU usage, as it drops the unsolicited packets before processing them.
        }
        \label{fig:cpu}
    \end{figure}

\begin{figure}[h]%{\columnwidth}
    \centering
        \includegraphics[width=0.5\columnwidth]{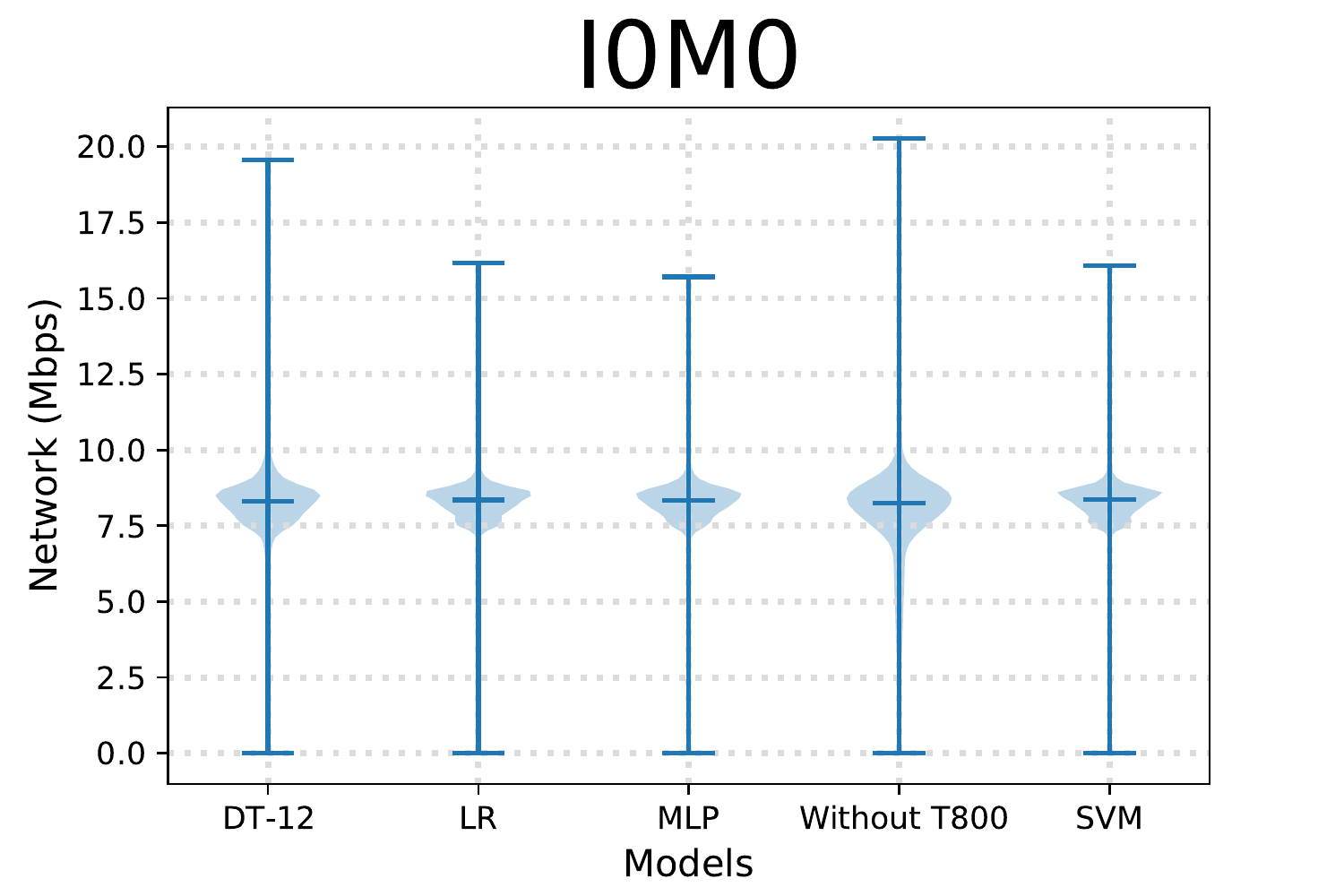}\hfill
        \includegraphics[width=0.5\columnwidth]{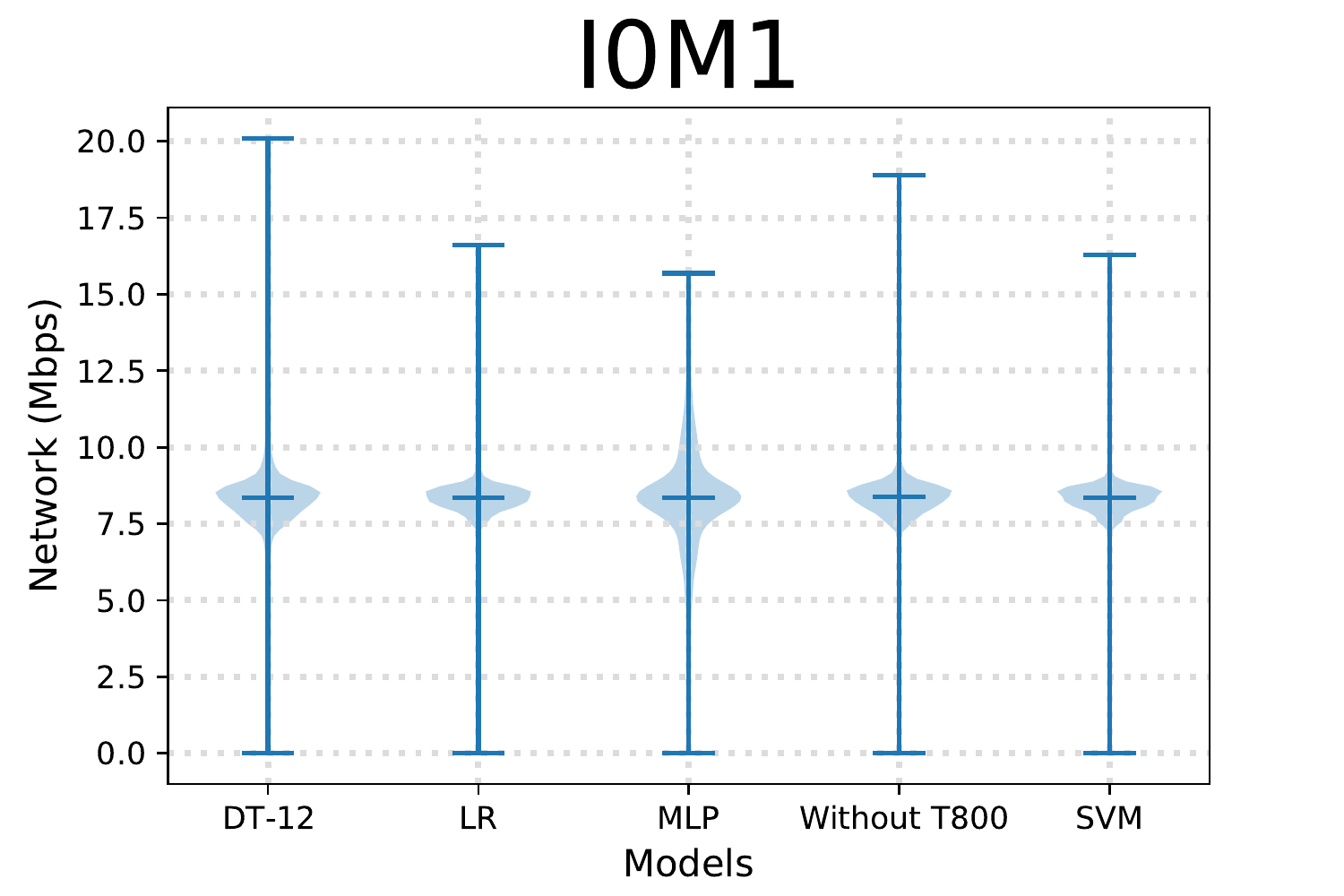}
        \includegraphics[width=0.5\columnwidth]{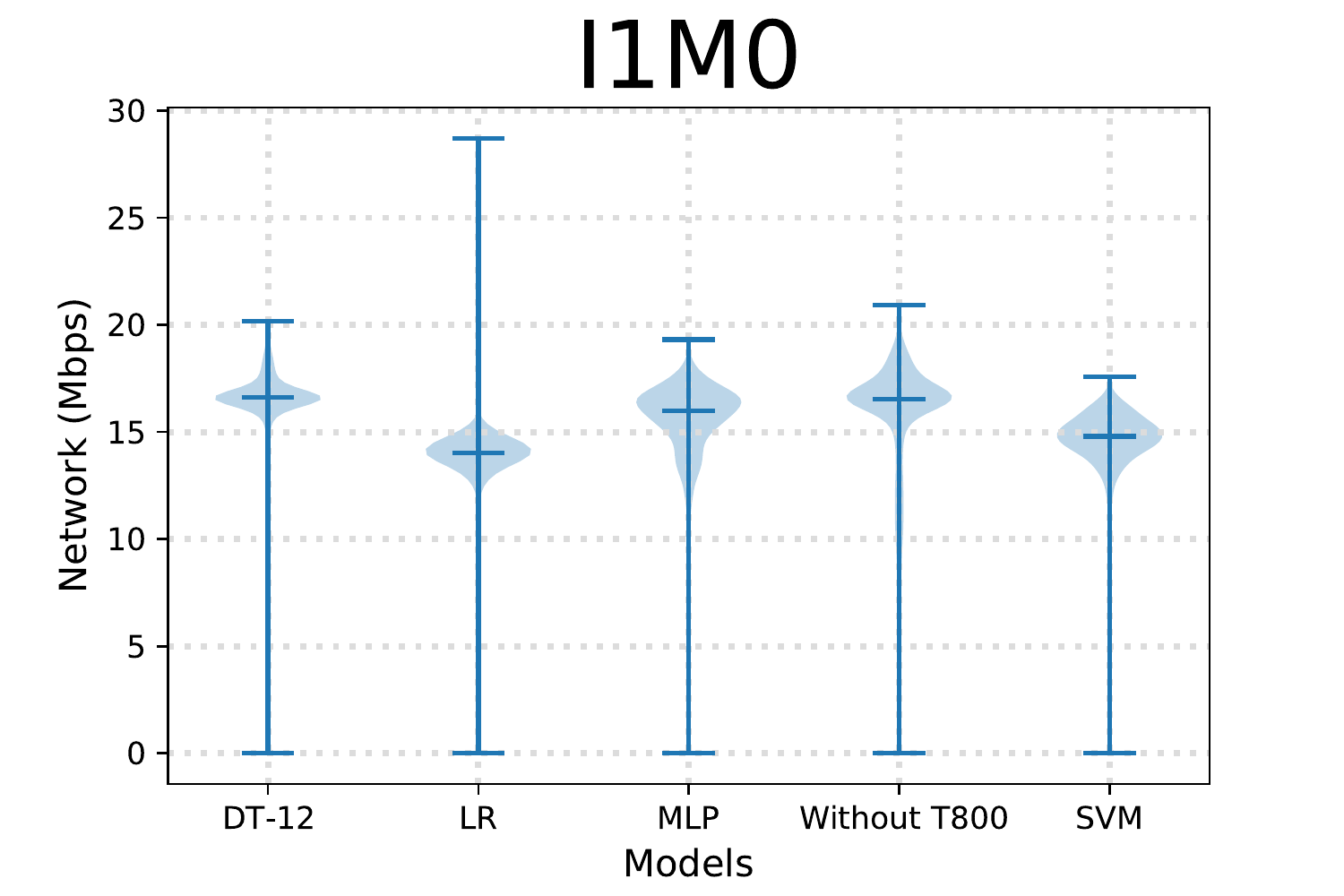}\hfill
        \includegraphics[width=0.5\columnwidth]{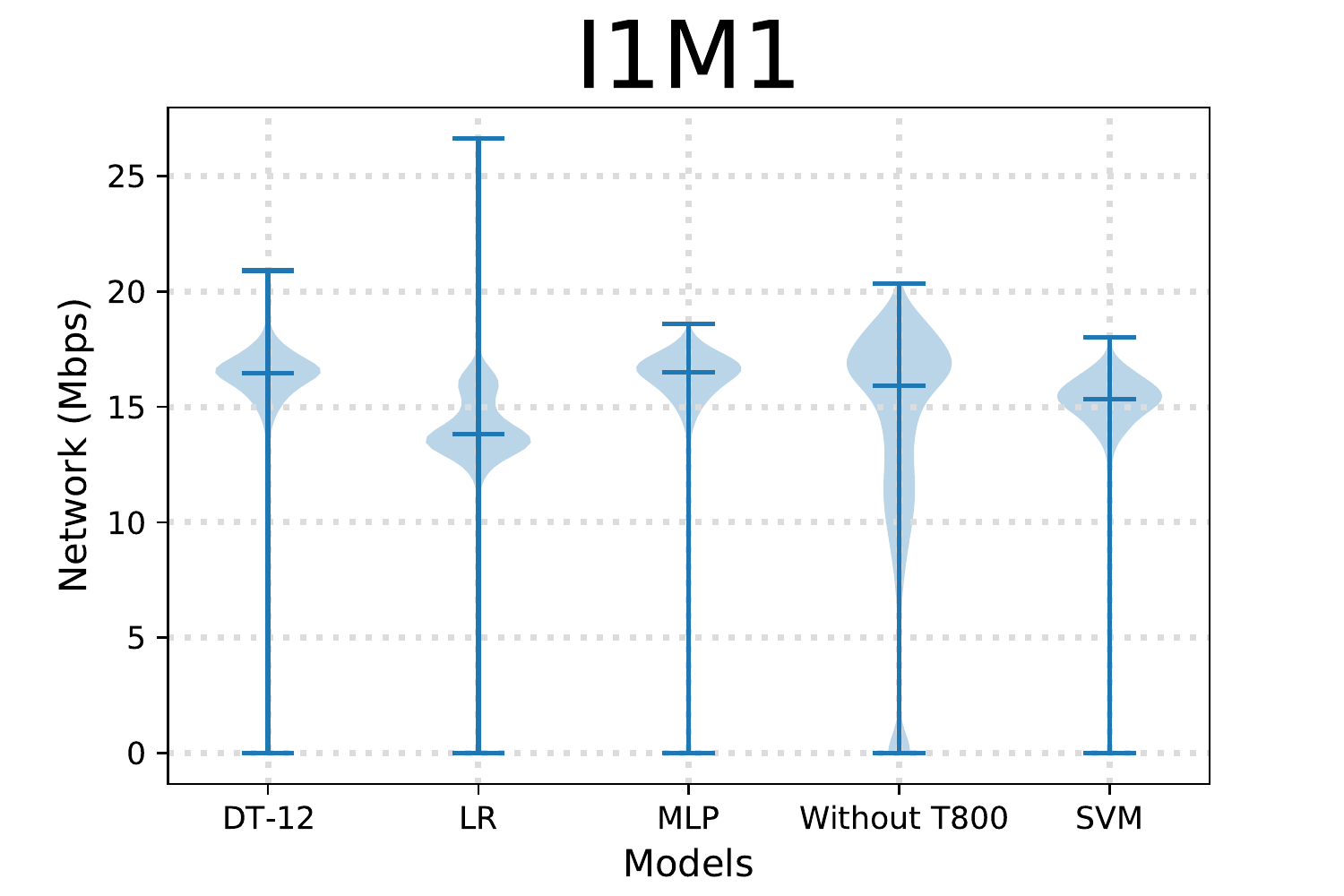}
        \caption{Consumption values obtained from experiments: Network Bandwidth. The row I0Mx corresponds to low-intensity traffic ($8$\,Mbps) and I1Mx to the high intensity ($16$\,Mbps).  The column IxM0 indicates the absence of malicious traffics, and IxM1 the presence.}
        \label{fig:rede}
    \end{figure}

\begin{figure}[h]%{\columnwidth}
    \centering
        \includegraphics[width=0.5\columnwidth]{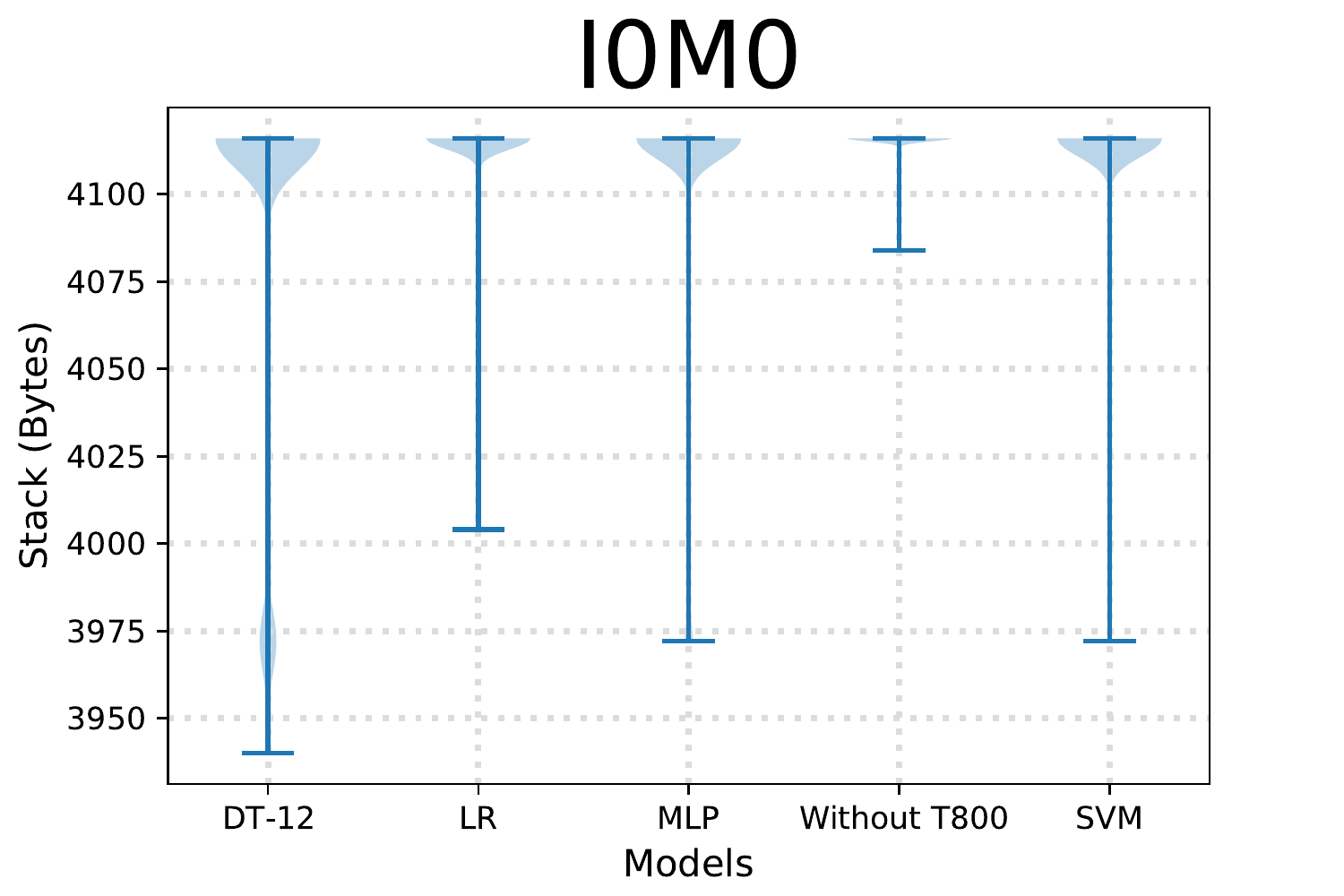}\hfill
        \includegraphics[width=0.5\columnwidth]{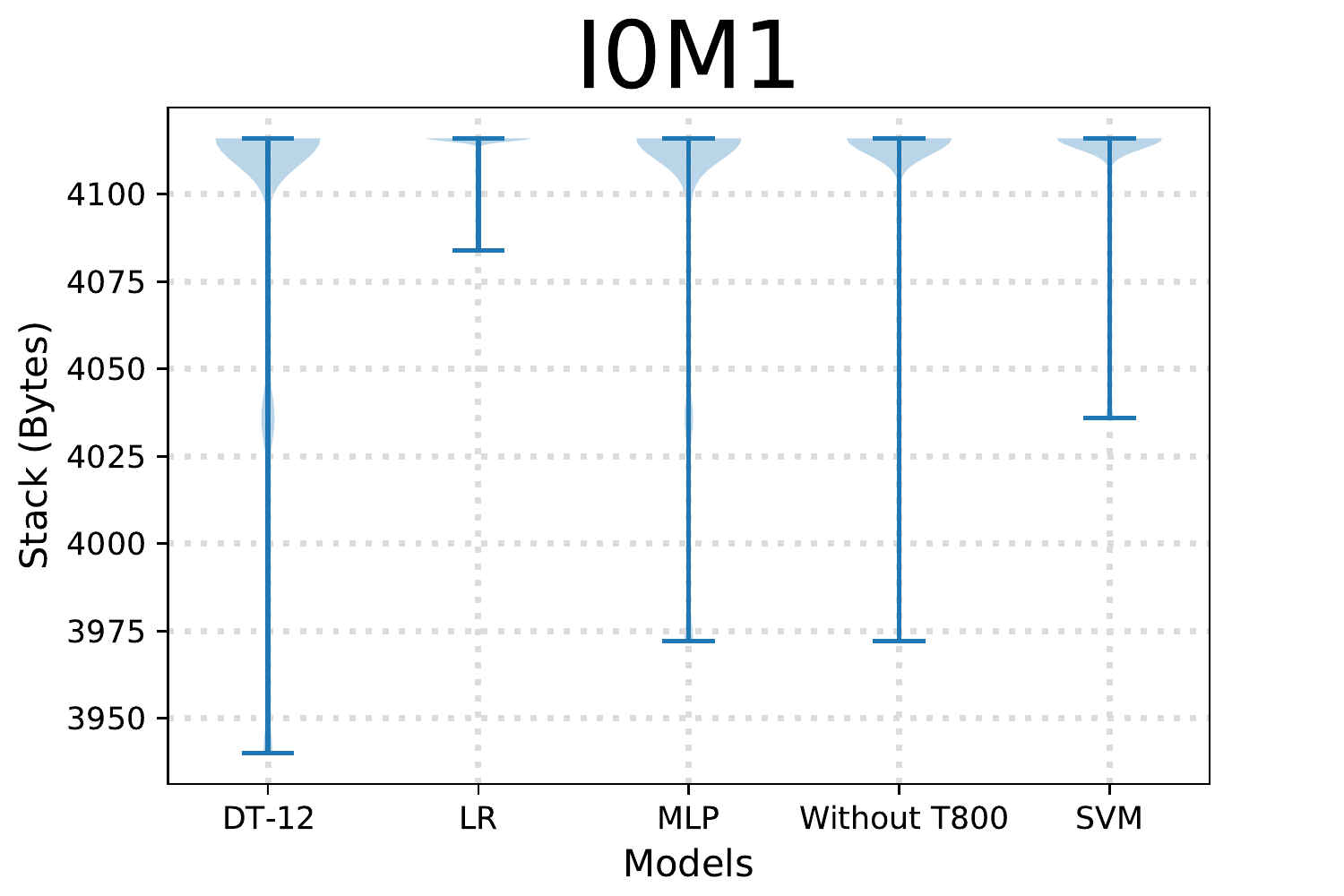}
        \includegraphics[width=0.5\columnwidth]{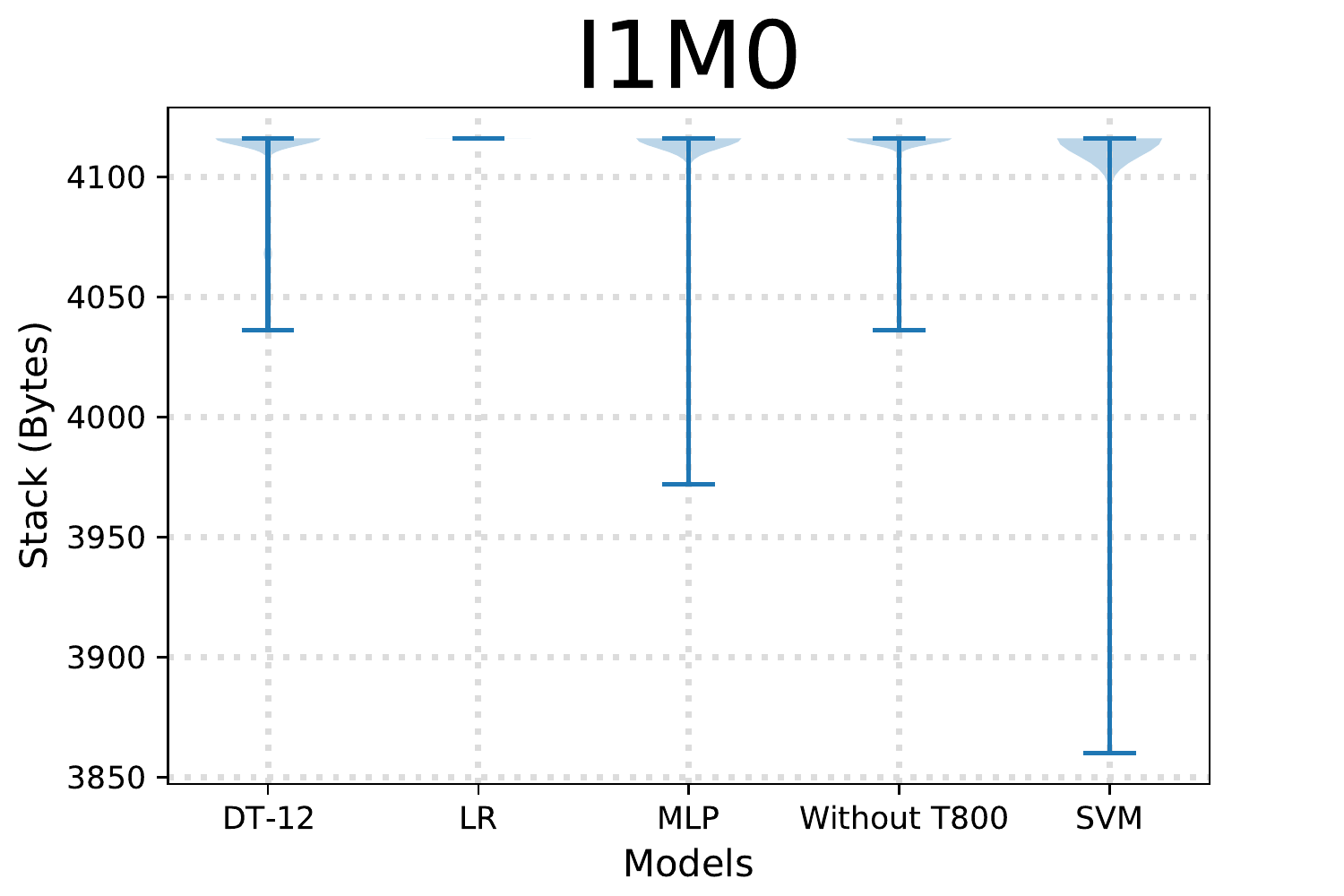}\hfill
        \includegraphics[width=0.5\columnwidth]{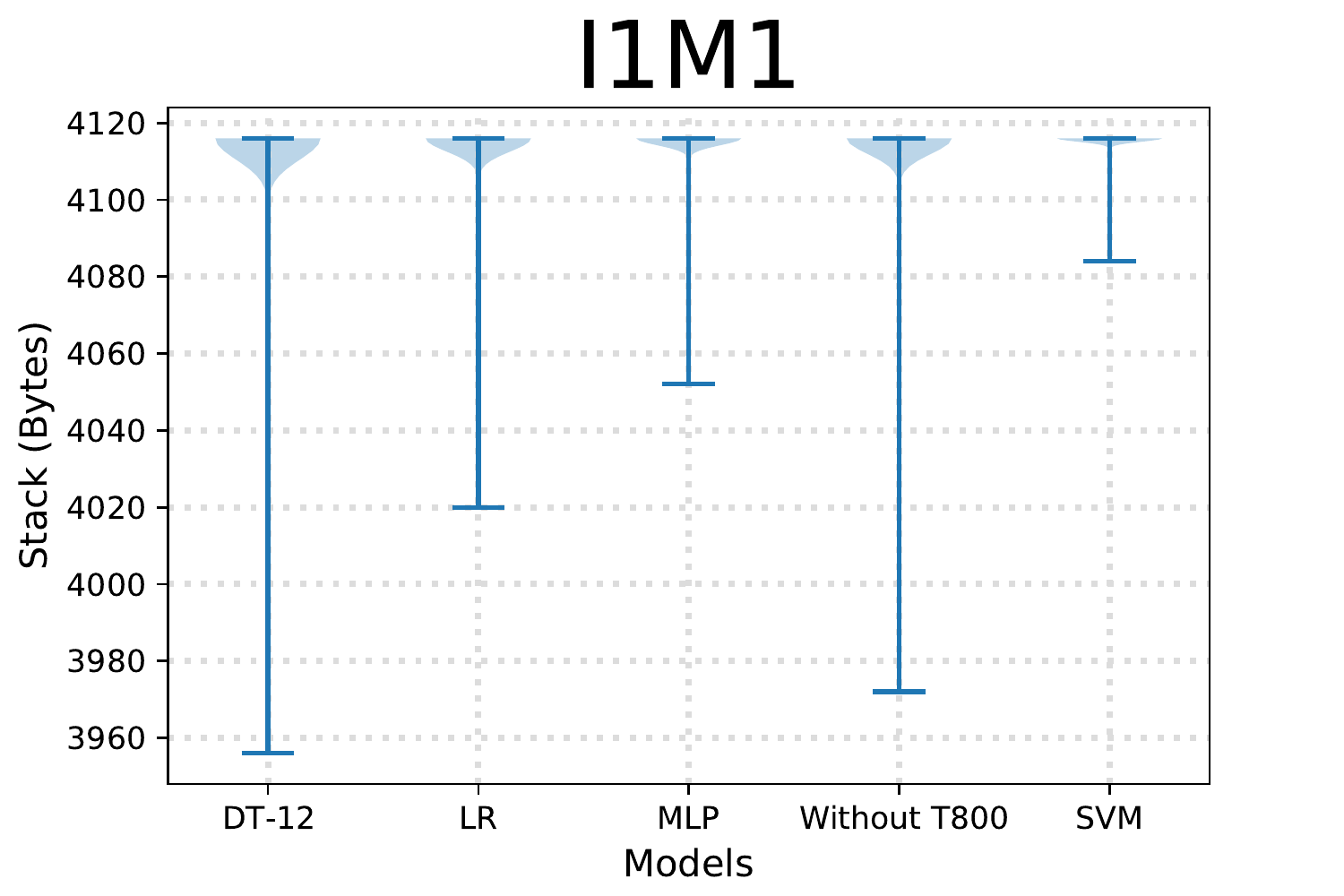}
        \caption{Consumption values obtained from experiments: Stack Memory Usage.  The memory footprint remains approximately the same compared to the system with the T800.  }
        \label{fig:stack}
    \end{figure}

\begin{figure}[h]%{\columnwidth}
    \centering
        \includegraphics[width=0.5\columnwidth]{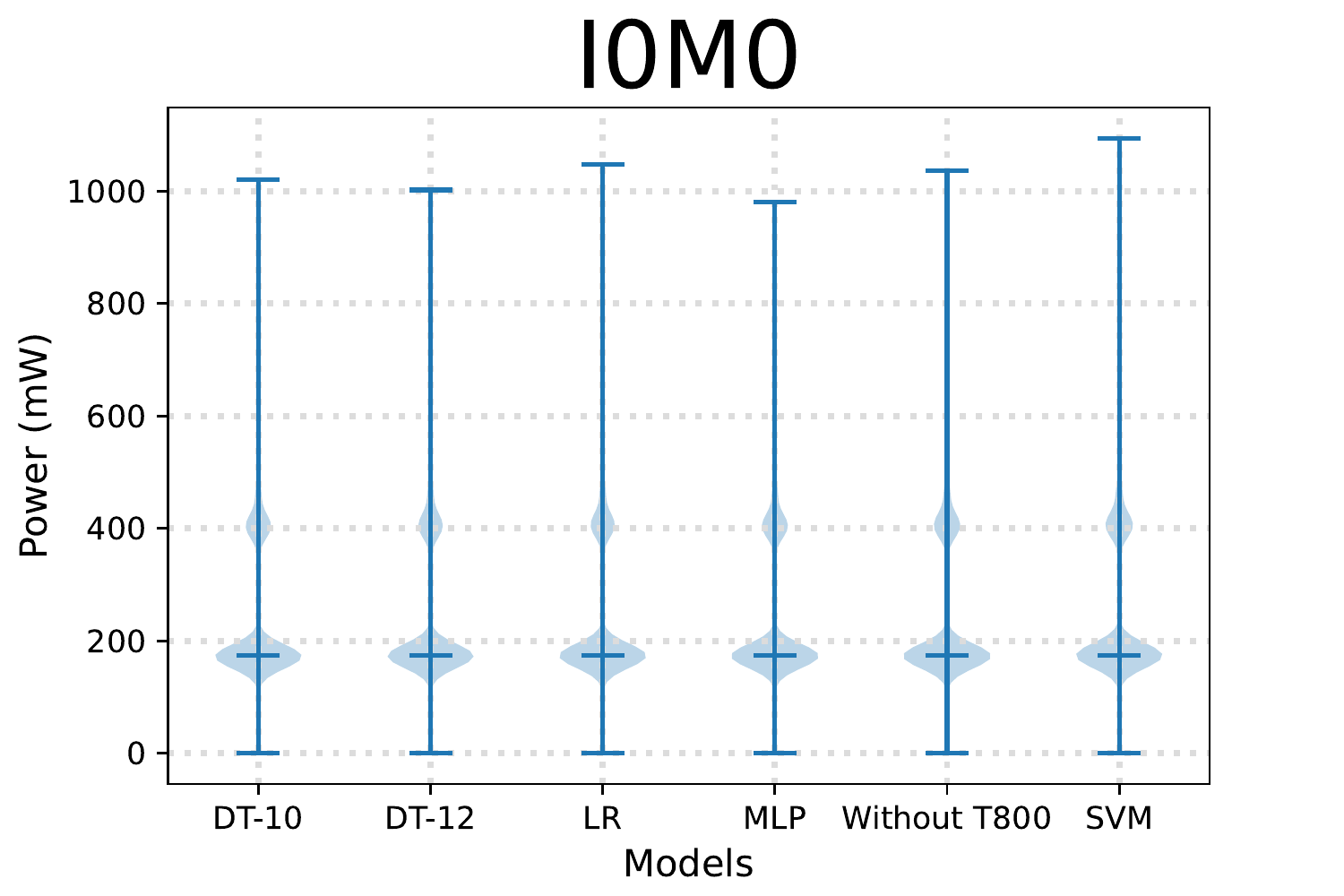}\hfill
        \includegraphics[width=0.5\columnwidth]{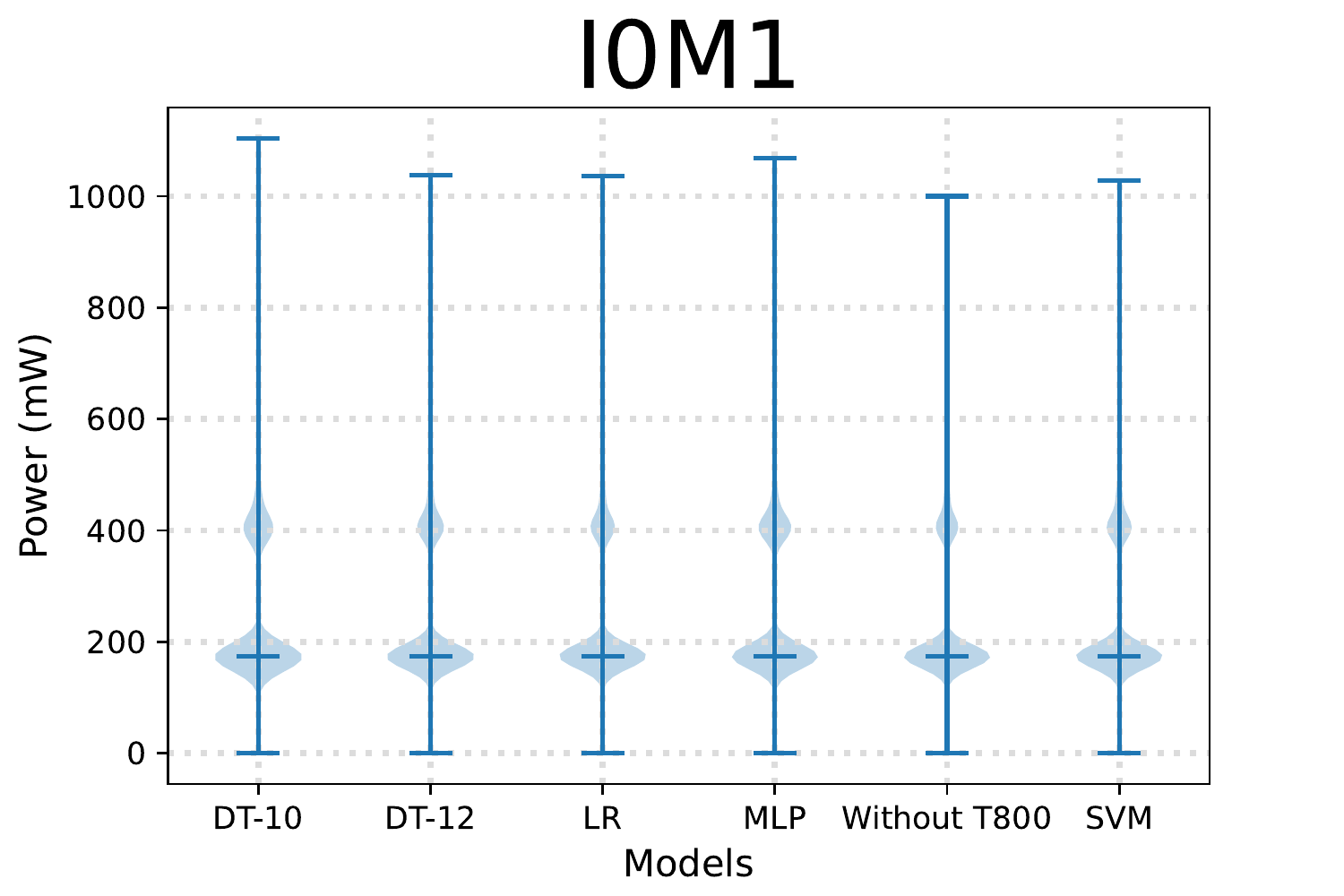}
        \includegraphics[width=0.5\columnwidth]{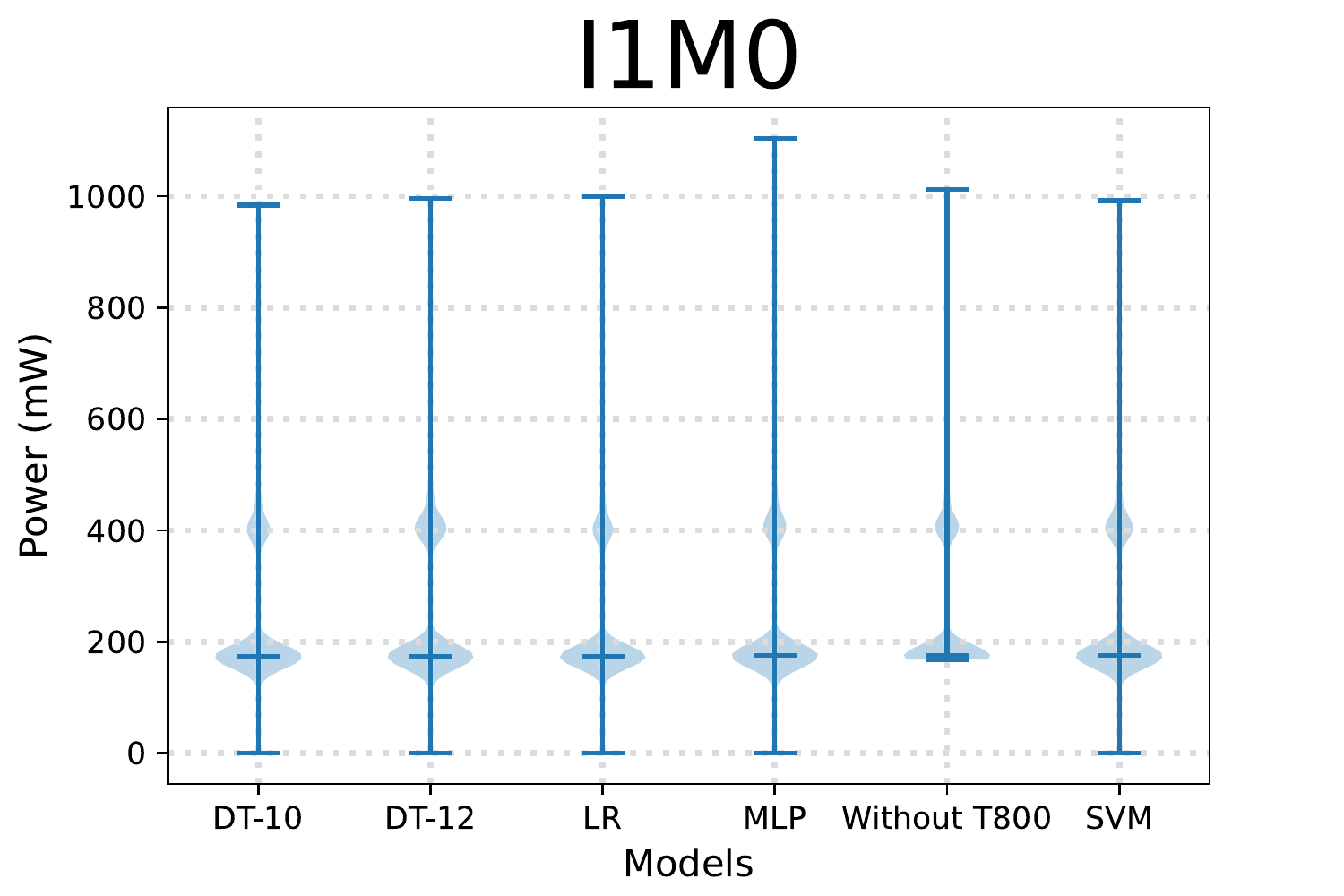}\hfill
        \includegraphics[width=0.5\columnwidth]{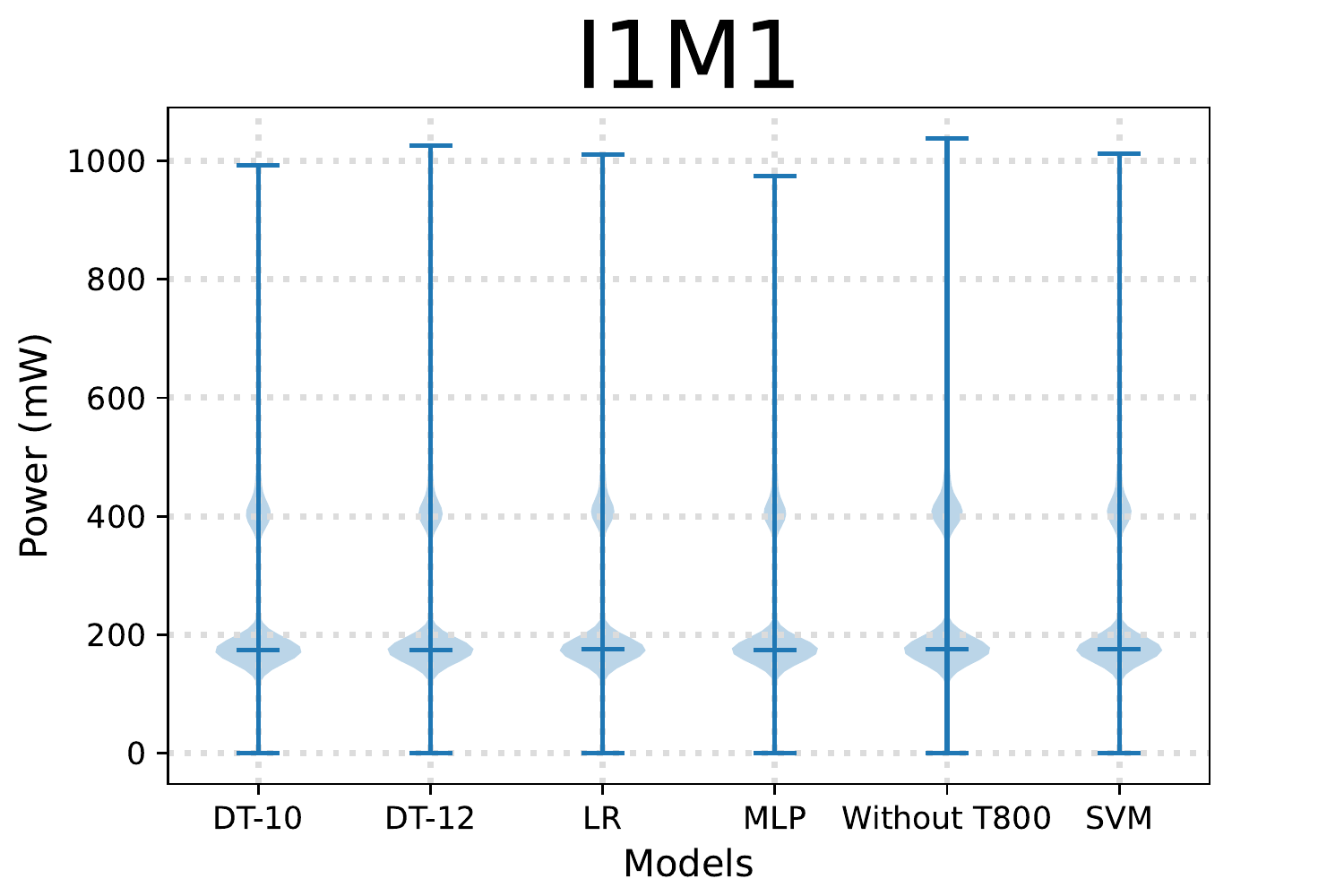}
        \caption{Consumption values obtained from experiments: Energy Usage.  The energy demands remain approximately the same compared to the system with the T800.}
        \label{fig:energia}
    \end{figure}

All process described in Section~\ref{sec:eval} results in the values presented in Figures~\ref{fig:cpu} to \ref{fig:energia}. These figures show the consumption of CPU, Network Bandwidth, Stack Memory usage, and Energy, respectively. We grouped metrics under interest in each of the corresponding experiment graphs. For instance, Figure~\ref{fig:cpu} presents CPU consumption for the I0M0 experiment with all models under consideration: decision tree (DT-12), logistic regression (LR), multilayer perceptron (MLP), SVM and without T800. We presented all the metrics as violin plots, meaning the samples represent a single aggregate value. However, the energy metric was continuously collected throughout the experiment, whereas we employed $3,600$ samples for the remaining metrics. Then, it was possible to analyze the median, maximum, and minimum. Finally, we compare the data from the executions of each model and those that do not use T800. The bars of each violin plot represent the extremities (maximum and minimum values) and the median (central bar) with the probability density function.

The CPU usage is present in the graphs of Figure~\ref{fig:cpu}. For the experiment I0M0, the medians of this metric are in the interval $[0.10, 0.12]$. For the experiment I0M1, the medians are in the interval $[0.10, 0.15]$. Finally, for I1M0, the medians are in the interval $[0.15, 0.19]$. Finally, for the experiment I1M1, the medians are in the interval $[0.16, 0.20]$. This data indicates a significant difference between the versions of T800 (all models) and those without T800. Furthermore, Figure~\ref{fig:rede} shows the network bandwidth. The I0 experimental levels represent a traffic intensity of $8$\,Mbps (median), and the I1 level exposes a median of $16$\,Mbps. Therefore, this metric's value remains close to the benign traffic intensity specified for the experiment. 

Moreover, the stack memory metric is showed in Figure~\ref{fig:stack}. The resulting medians remain constant with the value of $4116$ bytes, and the interquartile ranges show that the variation in this metric is negligible because of the magnitude being in tens of bytes. For last, Figure~\ref{fig:energia} displays the energy consumption. In them, most medians are in the range of $[174, 176]$.

% Portanto, através da análise de todos as métricas obtidas, constata-se que a utilização do T800 não impacta a desempenho do LwIP de nenhuma forma significativa.

    \begin{figure}[h]%{\columnwidth}
    \centering
        \includegraphics[width=0.5\columnwidth]{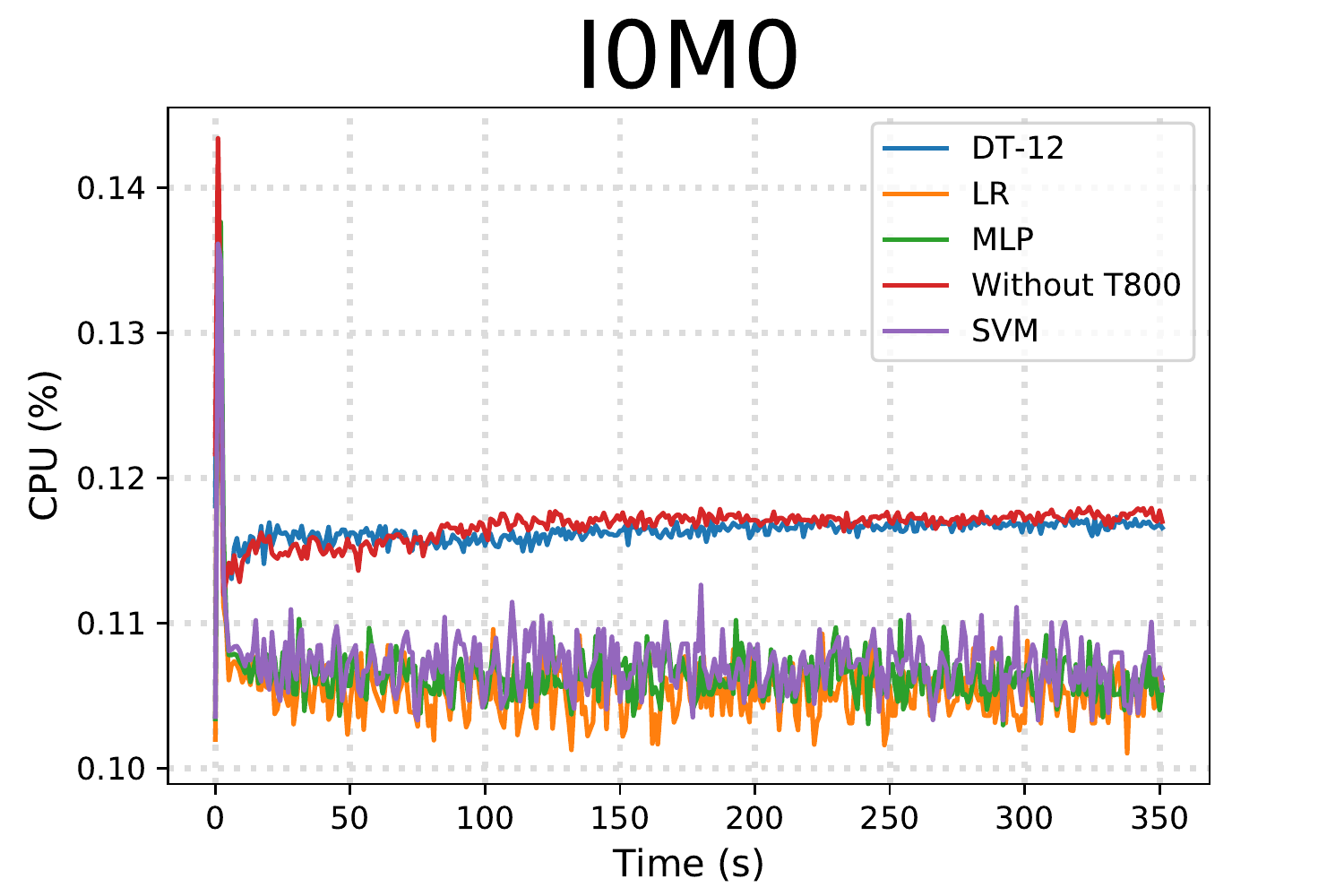}\hfill
        \includegraphics[width=0.5\columnwidth]{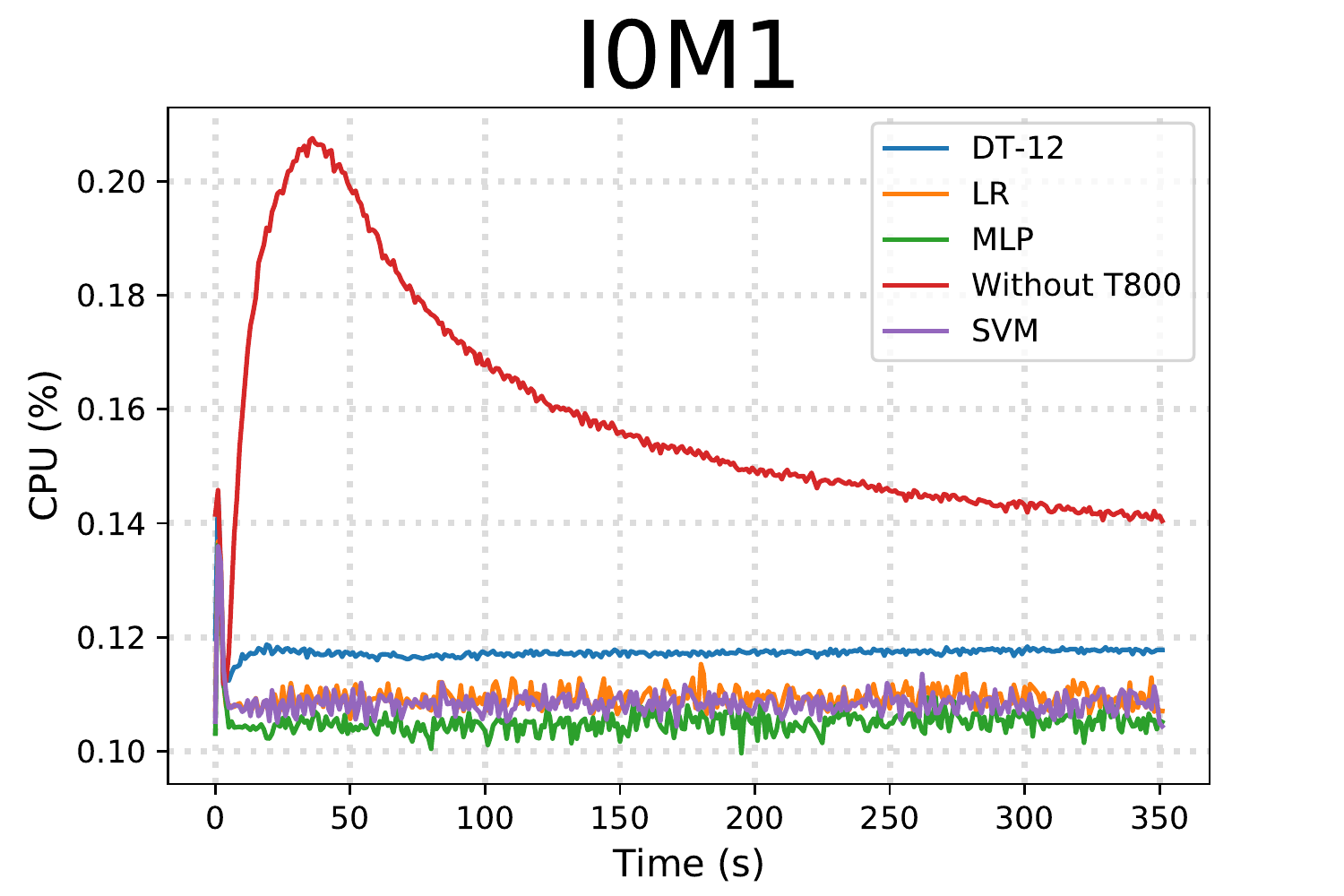}
        \includegraphics[width=0.5\columnwidth]{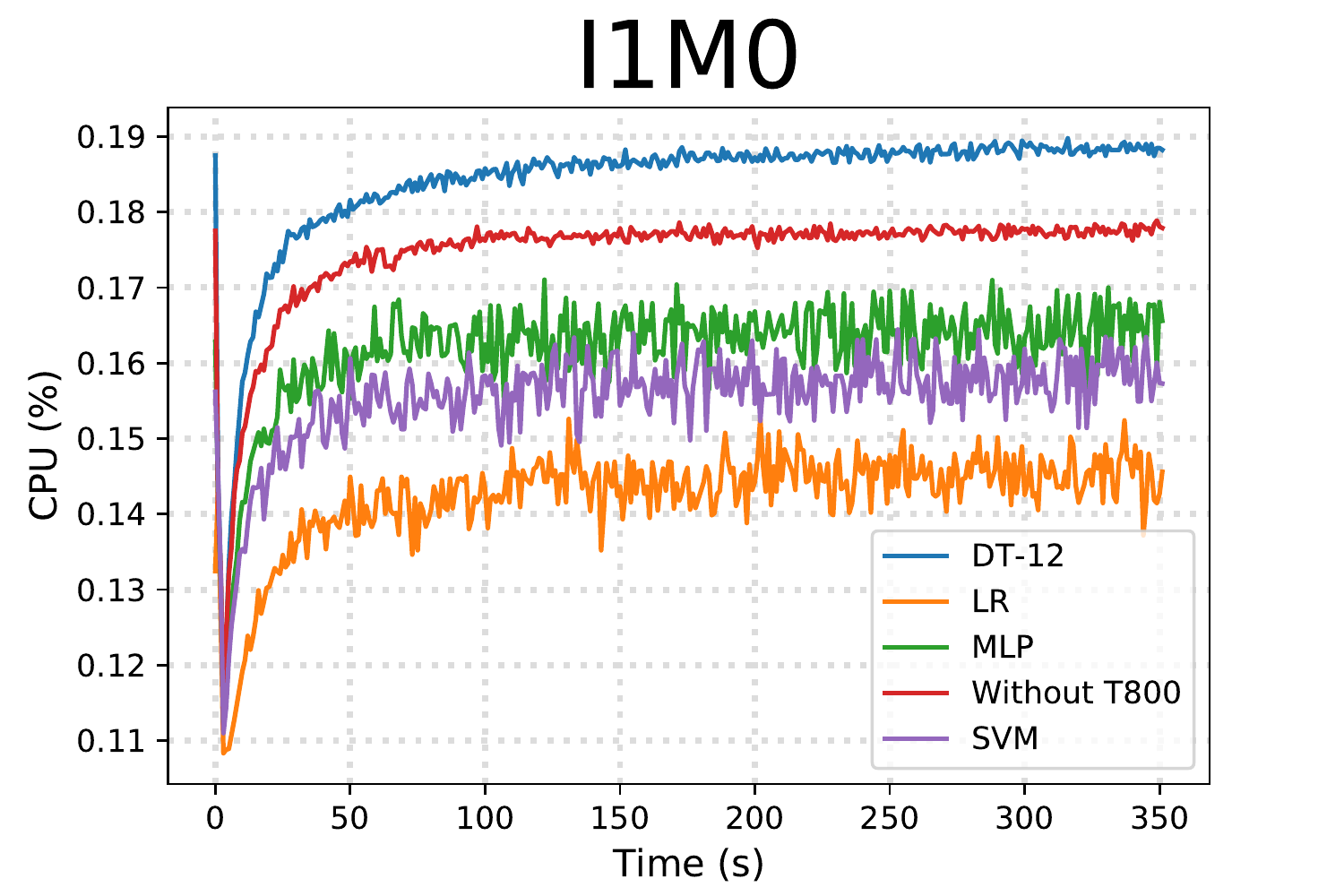}\hfill
        \includegraphics[width=0.5\columnwidth]{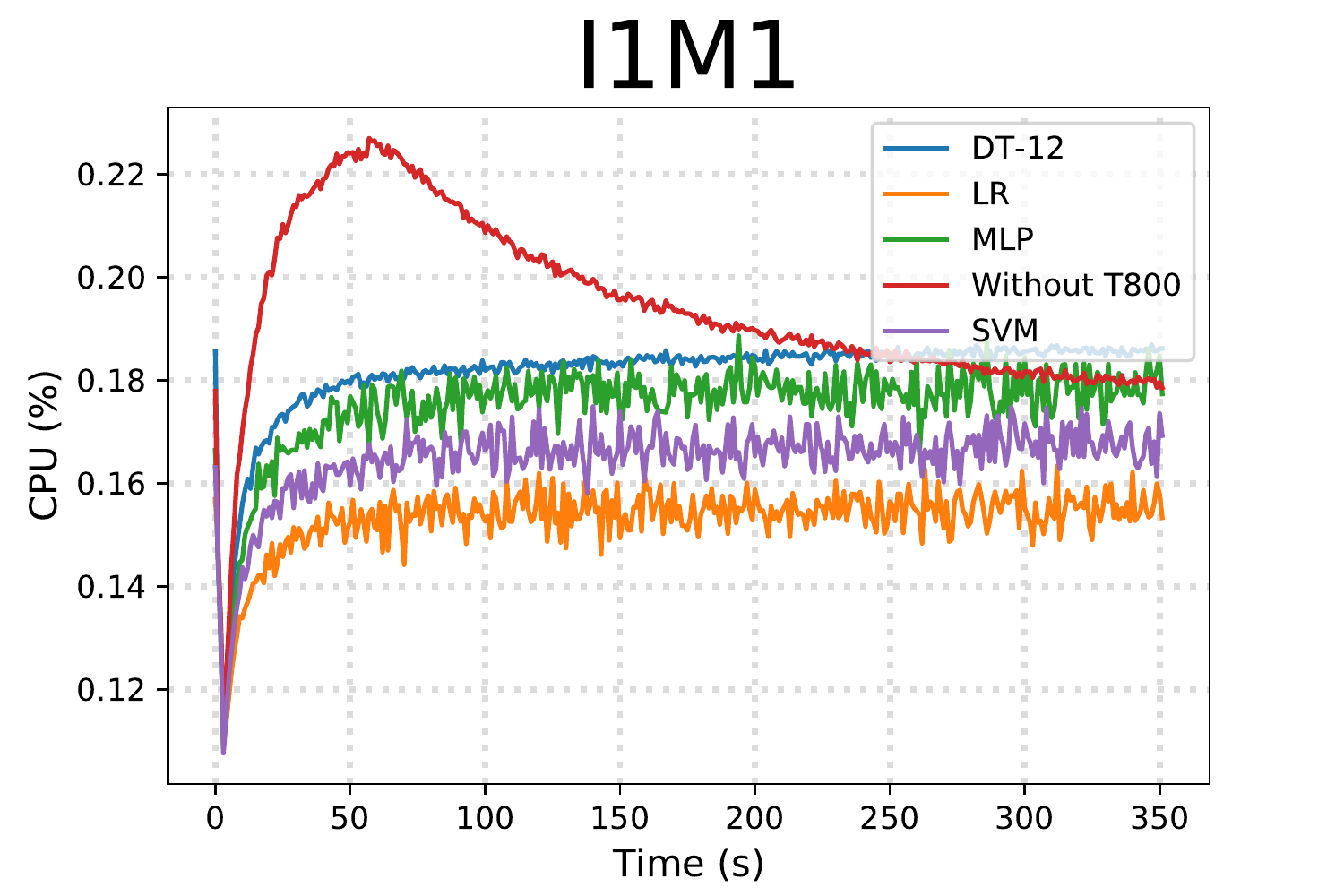}
        \caption{Time series of the CPU consumption values obtained through the experiments.  We observed that the experiments executed without the T800 demand more CPU than those with a packet filter.  For instance, the time series without T800 presents a high peak, and it is due to the burst of the malicious packets received by the network interface.  On the other hand, each machine learning model rejected those packets causing a steady-state behavior.}
        \label{fig:cpu_timeseries}
    \end{figure}

Besides, it is relevant to analyze the time series of the CPU usage shown in Figure~\ref{fig:cpu_timeseries}. The results of the experiments using machine learning models trained by Tensorflow presented lower computing consumption (SVM, Logistic Regression, and Multilayer Perceptron). It suggests that the framework optimizations make the models suitable for our low computing capacity systems context. Additionally, the graphs display an apparent discrepancy in performance when malicious traffic is present in the network flow. For instance, the baseline policy (AN - without T800) CPU usage is much higher than the filtering policies that use machine learning models in the I1M1 and I0M1 graphs. However, after the malicious traffic is interrupted, the CPU usage decays. Thus, T800 provides (\textit{i}) security protection against the reconnaissance (Cyber Kill Chain~\cite{ckc}) and (\textit{ii}) a reduction in the IoT system resource usage.

\subsection{Influence of Factors}

To analyze the influence of the selected factors on the variation of CPU usage values, we consider the presence of a machine learning based filtering policy (\textbf{A}) as a factor of the experiments. Thus, as presented in Table~\ref{tab:influence_of_factors_ed}, besides the factors and levels of the experiment design already presented in Section~\ref{sec:eval}, the values present (A1) or absent (A0) were introduced to make up the new factor \textbf{A}. In addition, the levels were mapped for discrete values so that $T0 = -1$ and $T1 = 1$ for any factor $A$. In this way, the experiment and such performance metric started to be represented by Table~\ref{tab:fatorial_completo_T},  where the variables $y_{i,j}$ are the mean of the values obtained in the replication $j$ of the experiment $i$. From that, it was realized that the regression $\mathbf{X}\mathbf{Q} = \bar{\mathbf{Y}}$ that considers a additive system model under analysis, as presented by equations \ref{eqX}, and \ref{eqQY}.

\begin{equation}\label{eqX}
\mathbf{X} = \begin{bmatrix}
\vdots & \vdots & \vdots & \vdots & \vdots & \vdots & \vdots & \vdots\\
1 & T & I & M & TI & TM & IM & TIM\\
\vdots & \vdots & \vdots & \vdots & \vdots & \vdots & \vdots & \vdots\\
\end{bmatrix} %\text{and} %\hspace{10mm}
\end{equation}

\begin{equation}\label{eqQY}
\mathbf{Q} = \begin{bmatrix}
q_0\\
q_T\\
q_I\\
q_M\\
q_{TI}\\
q_{TM}\\
q_{IM}\\
q_{TIM}
\end{bmatrix} \text{and}\,\,\,   
\bar{\mathbf{Y}} = \begin{bmatrix}
\bar{Y_1}\\
\bar{Y_2}\\
\bar{Y_3}\\
\bar{Y_4}\\
\bar{Y_5}\\
\bar{Y_6}\\
\bar{Y_7}\\
\bar{Y_8}
\end{bmatrix}
\end{equation}%\hspace{10mm}

\begin{table}[h]
    \caption{Properties of the traffic during the experiments.}
    \centering
    \resizebox{\columnwidth}{!}{%
    \begin{tabular}{lll}
    \toprule
    \textbf{Property} & \textbf{Level} & \textbf{Code}\\
    \midrule
    Machine learning policy & Absent or Present & A0, A1\\
    Traffic intensity & 8\,Mbps or 16\,Mbps & I0, I1\\
    Malicious traffic & Absent or Present & M0, M1\\
    \bottomrule
    \label{tab:influence_of_factors_ed}
\end{tabular}
}
\end{table}

\begin{table}[t]
    \centering
    \caption{Representation of the complete factorial planning with adding the factor T and discrete levels.}
    \resizebox{\columnwidth}{!}{%
    \begin{tabular}{ccrrrrrrrcccc}
        \toprule
        $\mathbf{i}$ & $\mathbf{1}$ & $\mathbf{T}$ & $\mathbf{I}$ & $\mathbf{M}$ & $\mathbf{TI}$ & $\mathbf{TM}$ & $\mathbf{IM}$ & $\mathbf{TIM}$ & $\mathbf{Y}_{i,1}$ & \ldots & $\mathbf{Y}_{i,175}$ & $\bar{\mathbf{Y}}_i$ \\
        \midrule
        $1$ & $1$ & $-1$ & $-1$ & $-1$ & $1$ & $1$ & $1$ & $-1$ & $y_{1,1}$ & \ldots & $y_{1,10}$ & $\hat{y_{1}}$\\
        $2$ & $1$ & $-1$ & $-1$ & $1$ & $1$ & $-1$ & $-1$ & $1$ & $y_{2,1}$ & \ldots & $y_{2,10}$ & $\hat{y_{2}}$\\
        $3$ & $1$ & $-1$ & $1$ & $-1$ & $-1$ & $1$ & $-1$ & $1$ & $y_{3,1}$ & \ldots & $y_{3,10}$ & $\hat{y_{3}}$\\
        $4$ & $1$ & $-1$ & $1$ & $1$ & $-1$ & $-1$ & $1$ & $-1$ & $y_{4,1}$ & \ldots & $y_{4,10}$ & $\hat{y_{4}}$\\
        $5$ & $1$ & $1$ & $-1$ & $-1$ & $-1$ & $-1$ & $1$ & $1$ & $y_{5,1}$ & \ldots & $y_{5,10}$ & $\hat{y_{5}}$\\
        $6$ & $1$ & $1$ & $-1$ & $1$ & $-1$ & $1$ & $-1$ & $-1$ & $y_{6,1}$ & \ldots & $y_{6,10}$ & $\hat{y_{6}}$\\
        $7$ & $1$ & $1$ & $1$ & $-1$ & $1$ & $-1$ & $-1$ & $-1$ & $y_{7,1}$ & \ldots & $y_{7,10}$ & $\hat{y_{7}}$\\
        $8$ & $1$ & $1$ & $1$ & $1$ & $1$ & $1$ & $1$ & $1$ & $y_{8,1}$ & \ldots & $y_{8,10}$ & $\hat{y_{8}}$\\
        \bottomrule
    \end{tabular}%
    }
    \label{tab:fatorial_completo_T}
\end{table}

Where $\mathbf{Q}$ represents the set of estimated parameters by the least squares algorithm, $\mathbf{X}$ the set of predictors related to each factor and the interactions between them, and $\bar{\mathbf{Y}}$ the sample mean values generated by the experiments. Thus, it is possible to not only obtain the variations caused by each factor and their combinations ($SS_A$, equation~\ref{eqSSA}) but also to compute the variation attributed to experimental errors ($SSE$, equation~\ref{eqSSE}) and a value corresponding to the total variation of the metric.

\begin{equation}\label{eqSSA}
SS_A = 2^3 \hspace{2mm} 10 \hspace{2mm} q_A
\end{equation}

\begin{equation}\label{eqSSE}
SSE = \sum_{i=1}^8 \sum_{j=1}^{10} (y_{ij} - \bar{y}_i)^2
\end{equation}
% \hspace{10mm}

Finally, we computed the influence factor as the variation caused by a factor divided by the total variation for each filtering policy. Then, it calculated an average of all the policies as a form of aggregation (Equation~\ref{eqSST}). 

\begin{equation}\label{eqSST}
\begin{split}
%\hspace{10mm}
SST = SS_T + SS_I + SS_M + SS_{TI} + SS_{TM} + SS_{IM} \\ 
+ SS_{TIM} + SSE
\end{split}
\end{equation}

The obtained results are in Table~\ref{tab:fatores}. Each column in the table quantifies directly how much the factor impacts CPU utilization.  As seen, the \textbf{I} factor (traffic intensity) impacts most in CPU utilization, $46\%$, $52\%$, $57\%$, $63\%$, and $72$\% for Linear Regression (LR), without T800 (w/o T800), Support Vector Machine (SVM), Multilayer Perceptron (MLP), and Decision Tree (DT-12), respectively.  For example, the traffic intensity influences from $46$\% to $72$\% of the current observed CPU  utilization.  The presence of T800 (factor \textbf{A}) is the second most impacting factor of CPU, followed by the incidence of malicious traffic (factor \textbf{M}).  On the other hand, the interaction of the factors (columns \textbf{AI}, \textbf{AM}, \textbf{IM}, and \textbf{AIM}) are negligible.  The last column (Err) is the error observed in the least square regression due to the systems' stochastic behavior.  Finally, the last row is the simple average of the values.  In conclusion, the traffic intensity (\textbf{I}) corresponds to $58$\% on average of the currently observed CPU utilization.  T800 (\textbf{A}) leads to an overhead of $12$\%.  The malicious traffic (\textbf{M}) yields $10$\%.  Furthermore, the interaction of the factors corresponds to $4$\%, and the observed error is $14$\%.

The influence of factors indicates that the predominant factor in the CPU usage value is the benign traffic produced by the \textit{IPerf} traffic generator (I). Also, the results obtained for the presence or absence of the machine learning models showed that this factor has little influence on the variation of CPU usage. Therefore, our results suggest that T800 in lightweight~IP incurs low computing overhead regardless of the network workload (high security and low traffic or one with low security and high traffic), being a feasible solution for intelligent packet filtering mechanism on microcontroller-based devices.

\begin{table}[h]
\centering
    \caption{Results of the influence of factors analysis for the CPU usage metric. Here, \textbf{A} represents the presence of a filtering policy in the T800, \textbf{I} represents the \textit{IPerf} network traffic intensity, and \textbf{M} represents the presence of malicious traffic in the network.}
    \resizebox{\columnwidth}{!}{%
    \begin{tabular}{ccccccccc}
        \toprule
        \textbf{Model} & \textbf{A} & \textbf{I} & \textbf{M} & \textbf{AI} & \textbf{AM} & \textbf{IM} & \textbf{AIM} & \textbf{Err}\\
        \midrule
        w/o T800         & 0.00 & 0.52 & 0.23 & 0.00 & 0.00 & 0.03 & 0.00 & 0.22\\
        MLP         & 0.12 & 0.63 & 0.07 & 0.01 & 0.03 & 0.00 & 0.02 & 0.12\\ 
        DT-12         & 0.03 & 0.72 & 0.05 & 0.02 & 0.06 & 0.01 & 0.01 & 0.11\\ 
        LR         & 0.27 & 0.46 & 0.09 & 0.00 & 0.03 & 0.00 & 0.01 & 0.12\\ 
        SVM         & 0.17 & 0.57 & 0.08 & 0.00 & 0.04 & 0.00 & 0.01 & 0.12\\ 
        \midrule
        Average & 0.12 & 0.58 & 0.10 & 0.00 & 0.03 & 0.00 & 0.01 & 0.14\\
        \bottomrule
    \end{tabular}%
    }
    \label{tab:fatores}
\end{table}

\section{Conclusion}\label{sec:conclusion}

This paper presented the T800 packet filter for Internet of Things (IoT) devices with low-computing power. The proposed architecture is adaptable to other platforms due to the instrumentation of the protocol stack used, simply identifying the packet interception point. It allows different filtering policies deployment through implementation in conjunction with the operating system. We show the solution's  effectiveness through the description of the implementation with the ESP32 development platform (FreeRTOS, lwIP TCP/IP stack, and ESP-IDF framework).

The experiments showed high efficiency in the chosen approach since the presence of the T800 reduced the consumption of computing resources in the presence of malicious traffic.  Moreover, the action of the packet filter allowed malicious traffic disposal and positively impacted the system.  The results suggest that our design is adequate for filtering network packets in IoT platforms with low impact.  Furthermore, our findings showed no statistically significant difference among the baseline, Decision Tree, and Multilayer Perception; therefore, the model with the highest accuracy is suitable for T800's deployment.

In future works, the objective is to implement \textit{stateful} packet filtering policies.  %---since the mechanism is implemented in the T800 but not tested in this paper. 
Another point of experimentation concerns an anomaly detection approach to detect never seen attacks~\cite{flids}. Finally, to integrate a \textit{zero-trust} based architecture aimed at IoT devices where the T800 serves as an enabler for such a solution. 

Finally, this paper represents an effort to bring more realism while approaching machine learning to devise security mechanisms in IoT with a microcontroller-based implementation. Understanding how to operate the algorithms (MLOps) in kernel mode is a challenge.  Moreover, evaluating how to translate data science notebooks analysis into an ESP32 implementation represents a turning point toward realistic testbeds.  All reproducible code used is in the repository \url{https://github.com/c2dc/T800}.

\section*{Acknowledgement}
This research had the support from the Programa de Pós-graduação em Aplicações Operacionais---PPGAO/ITA, from Fundação de Amparo à Pesquisa do Estado de São Paulo\,(FAPESP) granting $\#$2020/09850-0 and $\#$2021/09416-1, and from Conselho Nacional de Desenvolvimento Científico e Tecnológico\,(CNPq) granting $\#$157021/2021-1.

\bibliographystyle{elsarticle-num} 
\bibliography{references}


%\begin{thebibliography}{00}


%\bibitem{}

%\end{thebibliography}
\end{document}